\documentclass[%
 aip,
 amsmath,amssymb,
 reprint,%
]{revtex4-2}

\usepackage{graphicx}
\usepackage{dcolumn}
\usepackage{bm}

\usepackage{float}
\usepackage[utf8]{inputenc}
\usepackage[T1]{fontenc}
\usepackage{mathptmx}
\usepackage{etoolbox}
\usepackage{bbold}
\usepackage{color}
\usepackage{pgfplots}
\pgfplotsset{compat=newest}
\usepackage{tikz}
\usetikzlibrary{patterns}

\makeatletter
\def\@email#1#2{%
 \endgroup
 \patchcmd{\titleblock@produce}
  {\frontmatter@RRAPformat}
  {\frontmatter@RRAPformat{\produce@RRAP{*#1\href{mailto:#2}{#2}}}\frontmatter@RRAPformat}
  {}{}
}%
\makeatother

\makeatletter
\newcommand{\thickhline}{%
    \noalign {\ifnum 0=`}\fi \hrule height 1pt
    \futurelet \reserved@a \@xhline
}
\newcolumntype{"}{@{\hskip\tabcolsep\vrule width 1pt\hskip\tabcolsep}}
\makeatother

\begin{document}

\preprint{AIP/123-QED}

\title[Damped harmonic oscillator revisited: the fastest route to equilibrium]{Damped harmonic oscillator revisited: the fastest route to equilibrium}

\author{K. Lelas}
\email{klelas@ttf.unizg.hr}
\affiliation{Faculty of Textile Technology, University of Zagreb, Croatia}
\author{N. Poljak}
\affiliation{Department of Physics, Faculty of Science, University of Zagreb, Croatia}
\author{D. Juki\'c}
\affiliation{Faculty of Civil Engineering, University of Zagreb, Croatia}

\date{\today}

\begin{abstract}
Theoretically, solutions of the damped harmonic oscillator asymptotically approach equilibrium, i.e., the zero energy state, without ever reaching it exactly, and the critically damped solution approaches equilibrium faster than the underdamped or the overdamped solution. Experimentally, the systems described with this model reach equilibrium when the system's energy has dropped below some threshold corresponding to the energy resolution of the measuring apparatus. We show that one can (almost) always find an optimal underdamped solution that will reach this energy threshold sooner than all other underdamped solutions, as well as the critically damped solution, no matter how small this threshold is. We also comment on one exception to this for a particular type of initial conditions, when a specific overdamped solution reaches the equilibrium state sooner than all other solutions. We confirm some of our findings experimentally.
\end{abstract}

\maketitle

\section{Introduction}

Under which conditions does the damped harmonic oscillator return to the equilibrium state the fastest? Standard textbooks answer that the oscillator returns to the equilibrium state the fastest in the critically damped regime, regardless of the initial conditions \cite{Morin1,Pain,Berkeley,Morin2,Fluids}. Since none of the solutions to this model ever exactly reach the equilibrium state, but only asymptotically approach it, textbook authors usually justify this answer by considering the convergence speed of the general underdamped, critically damped and overdamped solutions towards the equilibrium state in the infinite time limit \cite{Morin1}. Indeed, in that asymptotic sense, the critically damped solution approaches the equilibrium state faster than the others. The situation becomes more subtle when experimental resolution is taken into account; one can determine the damping coefficient in the underdamped regime for which the oscillator settles down fastest to equilibrium by setting the maximal overshoot to be equal to the displacement resolution \cite{Armstrong}. Similarly, Heald\cite{Heald} sets the minimum detectable overshoot equal to the displacement resolution and computes the degree of underdamping that would be mistaken for critical due to the fastest return to the equilibrium. 

In this paper we theoretically find the time interval during which the envelope of the underdamped displacement is smaller in magnitude than the displacement of a critically damped oscillator using the Lambert W function \cite{Corless}. Depending on the damping coefficient, we determine the instant at which the envelope of underdamped displacement becomes larger than the displacement of the critically damped oscillator. We find that displacements at and after that moment are experimentally practically immeasurable for damping coefficients that are well within the underdamped regime, e.g.\,10\% smaller than the critical damping coefficient. 

In the ideal case, the equilibrium state is the zero energy state, with both the displacement and the velocity equal to zero, here we define the equilibrium state as the state in which the energy of the oscillator has dropped to a certain (infinitesimal) fraction of the initial energy, e.g.\,below a given experimental energy resolution. We show that it is possible to find an optimal damping coefficient in the underdamped regime for which the system's energy will fall below this predetermined threshold sooner than for any other underdamped, critically damped or overdamped oscillators with the same initial conditions. An exception to this is a special set of initial conditions for which an overdamped oscillator reaches the equilibrium state sooner than all others, regardless of whether we define the equilibrium state as a zero energy state or as a state with some infinitesimal fraction of initial energy. To check our statements experimentally, we designed an RLC circuit with variable parameters and measured the oscillating voltage across the resistor using a standard laboratory oscilloscope. The measurements are consistent with the theoretical findings.

\section{Short review of the model}   

The differential equation of the damped harmonic oscillator is of the form 
\begin{equation}
\ddot x(t)+2\gamma\dot x(t)+\omega_0^2x(t)=0\,,
\label{DHOeq}
\end{equation}
where $x(t)$ denotes the displacement from the equilibrium position as a function of time, the dots denote time derivatives, $\gamma>0$ is the damping coefficient and $\omega_0$ stands for the undamped oscillator angular frequency (the oscillator's natural frequency). The zero energy state of the system is achieved when both $x=0$ and $\dot x=0$. The form of the solution to the equation \eqref{DHOeq} depends on the relationship between $\gamma$ and $\omega_0$, producing three possible regimes. 

For $\gamma<\omega_0$ the system is in the \emph{underdamped regime} with the solution
\begin{equation}
\begin{split}
x_{ud}(t)=Ae^{-\gamma t}\cos(\omega t+\phi) \,, \hskip 3mm \textrm{with}
\\A=\sqrt{x_0^2+\frac{(v_0+\gamma x_0)^2}{\omega^2}}\hskip 3mm \textrm{and}\\
\phi=\arctan\left(-\frac{v_0+\gamma x_0}{\omega x_0}\right)\,.
\label{uds}
\end{split}
\end{equation}
Here, $\omega=\sqrt{\omega_0^2-\gamma^2}$ is the damped oscillator's angular frequency, $A$ and $\phi$ are constants determined from the initial conditions $x_0\equiv x(0)$ and $v_0\equiv \dot x(0)$. The displacement oscillates quasiperiodically within an exponentially decaying \emph{envelope} given by 
\begin{equation}
A(t)=A e^{-\gamma t}=\sqrt{x_0^2+\frac{(v_0+\gamma x_0)^2}{\omega^2}}e^{-\gamma t}\,.
\label{envelope}
\end{equation}
We want to emphasize here that the displacement maxima are not tangent to the envelope and it could be misleading to refer to the function \eqref{envelope} as the time dependent amplitude \cite{amplitude}. It is easy to check that the envelope of the velocity, $\dot x_{ud}(t)$, behaves similarly, and thus the system asymptotically approaches equilibrium.   

For $\gamma=\omega_0$ the system is in the \emph{critically damped regime} with the solution 
\begin{equation}
x_c(t)=\left(x_0+\left(v_0+\omega_0 x_0\right)t\right)e^{-\omega_0 t}.
\label{cds}
\end{equation} 
The solution approaches equilibrium without oscillating. 

For $\gamma>\omega_0$ the system is in the \emph{overdamped regime} with the solution 
\begin{equation}
x_{od}(t)=e^{-\gamma t}\left(\frac{v_0+(\gamma+\alpha)x_0}{2\alpha}e^{\alpha t}-\frac{v_0+(\gamma-\alpha)x_0}{2\alpha}e^{-\alpha t}\right),
\label{ods}
\end{equation}   
where $\alpha=\sqrt{\gamma^2-\omega_0^2}$. As in the critically damped regime, the solution approaches equilibrium without oscillating. 

The envelope of the underdamped oscillations decays as $e^{-\gamma t}$, and thus to compare the decays of the underdamped and critically damped solutions, we examine the ratio
\begin{equation}
\lim_{t \to \infty}\frac{x_c(t)}{e^{-\gamma t}}=\lim_{t \to \infty}\left(x_0+\left(v_0+\omega_0 x_0\right)t\right)e^{-(\omega_0-\gamma) t}=0,
\label{xc_xud}
\end{equation}
since $\omega_0 > \gamma$. Therefore, the critically damped solution approaches equilibrium faster than the underdamped solution. For the overdamped regime, the approach to the equilibrium in the infinite time limit is governed by the slowly decaying exponential $e^{(-\gamma+\alpha)t}$, and thus we evaluate 
\begin{equation}
\lim_{t \to \infty}\frac{x_c(t)}{e^{(-\gamma+\alpha) t}}=\lim_{t \to \infty}\left(x_0+\left(v_0+\omega_0 x_0\right)t\right)e^{-(\omega_0-\gamma+\alpha) t}=0,
\label{xc_xod}
\end{equation}
due to $\omega_0-\gamma+\alpha>0$. Again, the critically damped solution approaches the equilibrium faster than the overdamped one. The last statement is not completely general, since there is a possibility of a special case in which the initial displacement and velocity are of opposite sign and satisfy $|v_0|>|x_0\omega_0|$. In this case, it is possible to choose a damping coefficient $\gamma$ such that the factor multiplying the slowly decaying exponential is zero and only the quickly decaying exponential $e^{(-\gamma-\alpha)t}$ is present. In this case, the overdamped oscillator approaches the equilibrium state faster than the critically damped oscillator with the same initial conditions.  

We now consider the energy of the oscillator. If we multiply Eq. \eqref{DHOeq} by $\dot x(t)$, and rewrite the resulting first and third terms as time derivatives, we get  
\begin{equation}
\frac{\textrm{d}}{\textrm{d}t}\left(\frac{\dot x(t)^2}{2}+\frac{\omega_0^2x(t)^2}{2}\right)+2\gamma\dot x(t)^2=0\,.
\label{EvsD}
\end{equation}
If we consider, e.g., a mass on a spring in viscous fluid, the first term in the equation \eqref{EvsD} is the time derivative of the total mechanical energy (kinetic plus potential) per unit mass and the second term is the power of the dissipative force per unit mass. The ratio of the system's mechanical energy at time $t$ to its initial mechanical energy is
\begin{equation}
\frac{E(t)}{E_0}=\frac{\dot x(t)^2+\omega_0^2x(t)^2}{v_0^2+\omega_0^2x_0^2}\,.
\label{E}
\end{equation}
Using \eqref{E}, we can rewrite \eqref{EvsD} as a fractional rate of energy decrease
\begin{equation}
\frac{\textrm{d}}{\textrm{d}t}\left(\frac{E(t)}{E_0}\right)=-\frac{4\gamma}{v_0^2+\omega_0^2x_0^2}\dot x(t)^2\,.
\label{Eloss}
\end{equation}
Here, for later convenience, we define the ratio of the kinetic to total energy at any given time $t$ as
\begin{equation}
\beta(t)=\frac{\dot x(t)^2}{\dot x(t)^2+\omega_0^2x(t)^2}\,.
\label{beta}
\end{equation}
Note that, in the underdamped case, when $\beta_{ud}(t)=1$, the system is in the equilibrium position and possesses only kinetic energy. Conversely, when $\beta_{ud}(t)=0$ the system is at a turning point, and possesses only potential energy.  Using \eqref{beta}, we can rewrite \eqref{Eloss} using the natural logarithm as
\begin{equation}
\frac{\textrm{d}}{\textrm{d}t}\left(\ln\frac{E(t)}{E_0}\right)=-4\gamma\beta(t)\,,
\label{lossln}
\end{equation}
since the total energy is strictly real and positive.
Equation \eqref{lossln} is valid for all three oscillating regimes, i.e., for underdamped, critically damped and overdamped regimes. We find it especially useful when analyzing the underdamped case on account of its quasiperiodic motion. 

Eqs. \eqref{Eloss} and \eqref{lossln} are clearly related, but one must be aware of their implicit differences. For example, both loss rates have maxima, equal to zero, at moments when $\dot x(t)^2=0$ and $\beta(t)=0$, but the minima occur at different times. Compared to the undamped system, the underdamped oscillator achieves its maximum speed somewhat before it reaches the equilibrium position \cite{Peter} where $\beta(t) = 1$ . To avoid confusion, in what follows, we will refer to Eq. \eqref{lossln} as the \emph{energy loss rate}.   
  
To summarize, the solutions of the damped harmonic oscillator asymptotically approach the equilibrium state, but never exactly reach it. In that asymptotic sense, the critically damped oscillator returns to equilibrium the most quickly in all cases (apart from a specific case of an overdamped oscillator). In nature and in experiments, systems described by this model reach an effective equilibrium state in which the energy of the system has decreased to the level of the surrounding noise, or the energy resolution of the measuring apparatus. Following this line of thought, we will define a system to be in equilibrium for times $t>\tau$ such that 
\begin{equation}
\frac{E(\tau)}{E_0}=10^{-\delta}\,,
\label{tau}
\end{equation}
where $\delta>0$ is a dimensionless parameter that defines what fraction of the initial energy is left in the system. In the next section we prove that, for any finite value of $\delta$, one can always find an optimal solution in the underdamped regime ($\gamma<\omega_0$) which will reach equilibrium, in the sense of \eqref{tau}, sooner than all other underdamped or critically damped solutions with the same initial conditions. 

\section{The fastest route to equilibrium}  
\label{section:x0}
\subsection{Displacement analysis}
\label{subsection:displacement} 

\begin{figure}[h!t!]
\begin{center}
\begin{tikzpicture}
        \begin{axis}[
        width=0.485\textwidth,
        height=0.3\textwidth,
        xmin = 0,
        xmax = 5,
        ymin = -0.2,
        ymax = 1.4,
        xtick={0,0.5,...,4.5,5},
        ytick={-0.2,0,...,1.2,1.4},
        every tick label/.append style={font=\small},
        ylabel near ticks,
        xlabel near ticks,
        xlabel = {\small $t\,[\omega_0^{-1}]$},
        ylabel = {\small $x(t)/x_0$},
        legend entries = {\footnotesize $A(t),\quad \gamma=0.6\omega_0$ \\\footnotesize $x_{ud}(t),\quad \gamma=0.6\omega_0$\\\footnotesize $x_{c}(t), \quad\gamma=\omega_0$\\}
        ]
        \addplot [domain=0:5,samples=100,thick, red](\x,{1.25*exp(-0.6*\x)});
        \addplot [domain=0:5,samples=100,line width=0.25mm,densely dotted, blue](\x,{1.25*exp(-0.6*\x)*cos(deg(0.8*\x-0.6435))});
        \addplot [domain=0:5,samples=100,thick,dashed, black](\x,{(1+\x)*exp(-\x)});
        \addplot [domain=0:5,samples=100,line width=0.25mm,densely dotted, black!60!white](\x,{0.1});
        \addplot [domain=0:5,samples=100,line width=0.25mm,densely dotted, black!60!white](\x,{-0.1});
        \draw[<->,>=stealth] (axis cs: 1.5,0.1) -- (axis cs: 1.5,-0.1);
        \node[] at (axis cs: 1.8,-0.02) {\small $2|\Delta x|$};
        
        \draw[->,>=stealth] (axis cs: 0.6,0.888) -- (axis cs: 0.6,-0.2);
        \node[] at (axis cs: 0.73,-0.04) {\small $\tau_1$};
        \draw[->,>=stealth] (axis cs: 2.74,0.244) -- (axis cs: 2.74,-0.2);
        \node[] at (axis cs: 2.6,-0.04) {\small $\tau_2$};
        \end{axis}       
    \end{tikzpicture}
\end{center}
\vskip -5mm
\caption{The envelope $A(t)$ (solid red curve) and the complete underdamped solution $x_{ud}(t)$ with $\gamma= 0.6\omega_0$ (dotted blue curve) compared to the critically damped solution $x_c(t)$ (dashed black curve). The arrows denote the moments in time $\tau_1$ and $\tau_2$ when $A(t) = x_c(t)$ .
During the time interval $t\in(\tau_1,\tau_2)$ the envelope of the underdamped solution $x_{ud}(t)$ is closer to the equilibrium than $x_c(t)$. The magnitude of the displacement in the underdamped case becomes confined within the band $\pm 0.1 x_0$ before the critically damped case and remains within that region for all subsequent times.}  
\label{fig:Fig1}
\end{figure}
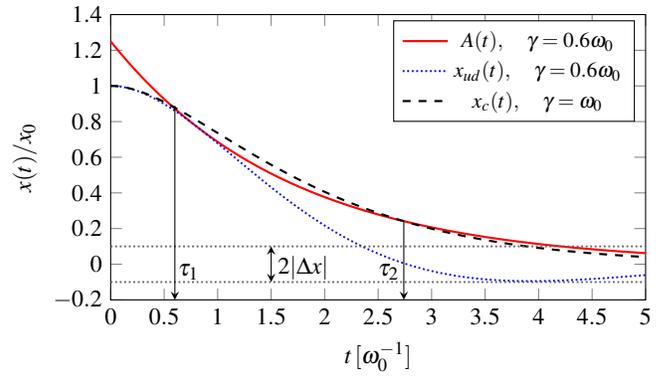

We focus on the initial conditions given by $x_0 > 0$ and $v_0=0$. The underdamped, critically damped, and overdamped solutions for this set of initial conditions are:
\begin{equation}
\begin{split}
x_{ud}(t)=A(t)\cos\left(\omega t+\phi\right)\,,\hskip 3mm \textrm{with}
\\A(t)=x_0\frac{\omega_0}{\omega}e^{-\gamma t}\, \hskip 3mm \textrm{and}\\
\phi=\arctan\left(-\frac{\gamma}{\omega}\right) \,,
\end{split}
\label{uds1}
\end{equation}
\begin{equation}
x_c(t)=x_0\left(1+\omega_0 t\right)e^{-\omega_0 t},
\label{cs1}
\end{equation} 
\begin{equation}
x_{od}(t)=\frac{x_0}{2\alpha}e^{-\gamma t}\left[(\gamma+\alpha)e^{\alpha t}-(\gamma-\alpha)e^{-\alpha t}\right].
\label{ods1}
\end{equation}   
The critical and overdamped solutions \eqref{cs1} and \eqref{ods1} have no zeros, and the overdamped solution approaches the equilibrium state more slowly. Thus, we can conclude that the overdamped solution is further from equilibrium than the critically damped solution, for any $t>0$ (we rigorously prove that $x_{od}(t)>x_c(t)$, for any $t>0$ in supplementary material \cite{SupplMat}, section SI). Therefore, we exclude the overdamped regime in what follows. In Fig.\ \ref{fig:Fig1} we show the envelope $A(t)$, the underdamped solution $x_{ud}(t)$ with $\gamma= 0.6\omega_0$, and the critically damped solution $x_c(t)$. Let us assume that Fig.\ \ref{fig:Fig1} models a real-life experiment in which the displacement resolution is, e.g., $\Delta x=\pm 0.1x_0$ (which would be a poor resolution, we take it here just to set an example).  From the point of view of that experiment, the underdamped solution settles down to equilibrium sooner than the critically damped solution. By setting the maximal overshoot equal to the displacement resolution, one can determine the damping coefficient in the underdamped regime for which the damped oscillator reaches equilibrium, within experimental resolution, as quickly as possible, as Armstrong\cite{Armstrong} has already shown. Here, we find the time interval $t\in(\tau_1,\tau_2)$ during which $A(t) < x_c(t)$ so that the envelope of the underdamped solution $x_{ud}(t)$ is actually \emph{closer} to the equilibrium than $x_c(t)$. 

We can gain analytical insight into the relationship between the underdamped and critically damped solutions by analyzing the dependence of the two crossing times, $\tau_1$ and $\tau_2$, on the damping coefficient. We equate $A(\tau_{1,2})=x_c(\tau_{1,2})$, and get 
\begin{equation}
e^{(\omega_0-\gamma)\tau_{1,2}}-\omega\tau_{1,2}-\frac{\omega}{\omega_0}=0\,,
\label{eqtau}
\end{equation}
with two solutions (see appendix \ref{AppendixLambert})
\begin{equation}
\tau_{1,2}=-\frac{1}{\omega_0-\gamma}W_{0,-1}\left(-\frac{\omega_0-\gamma}{\omega e^{\frac{\omega_0-\gamma}{\omega_0}}}\right)-\frac{1}{\omega_0},
\label{tau12}
\end{equation}
where $W_{0}(y)$ and $W_{-1}(y)$ are two different branches of the Lambert W function \cite{Corless}.

\begin{figure}[h!t!]
\begin{center}
\begin{tikzpicture}
    \begin{axis}[
        width=0.485\textwidth,
        height=0.3\textwidth,
        xmin = 0,
        xmax = 1.02,
        ymin = 0,
        ymax = 60,
        xtick={0,0.1,0.2,0.3,0.4,0.5,0.6,0.7,0.8,0.9,1.0},
        ytick={0,10,...,50,60},
        every tick label/.append style={font=\small},
        ylabel near ticks,
        xlabel near ticks,
        xlabel = {\small $\gamma\,[\omega_0]$},
        ylabel = {\small $\tau_{1,2}\,[\omega_0^{-1}]$},
        legend pos=north west
    ]
    \addplot [only marks, mark=o, sharp plot, red] table {
    0.05 0.043
    0.1 0.083
    0.15 0.121
    0.2 0.16
    0.25 0.199
    0.3 0.24
    0.35 0.284
    0.4 0.33
    0.45 0.38
    0.5 0.436
    0.55 0.498
    0.6 0.57
    0.65 0.655
    0.7 0.758
    0.75 0.887
    0.8 1.06
    0.85 1.31
    0.9 1.73
    0.95 2.66
    };
    \addplot [only marks, sharp plot, mark=x] table {
    0.05 0.062
    0.1 0.14
    0.15 0.234
    0.2 0.34
    0.25 0.477
    0.3 0.635
    0.35 0.825
    0.4 1.056
    0.45 1.34
    0.5 1.695
    0.55 2.148
    0.6 2.74
    0.65 3.535
    0.7 4.65
    0.75 6.3
    0.8 8.917
    0.85 13.6
    0.9 23.81
    0.95 58.41
    };
    \legend{$\tau_1$,$\tau_2$}
    \end{axis}       
\end{tikzpicture}
\end{center}
\vskip -5mm
\caption{The solutions to equation \eqref{tau12} for values of $\gamma[\omega_0]$ from 0.05 to 0.95 in steps of 0.05.  Between times $\tau_1$ and $\tau_2$, the underdamped solution $x_{ud}(t)$ oscillates with an envelope $A(t)$ that is smaller in magnitude than the critically damped solution $x_c(t)$.}
\label{fig:Fig2}
\end{figure}
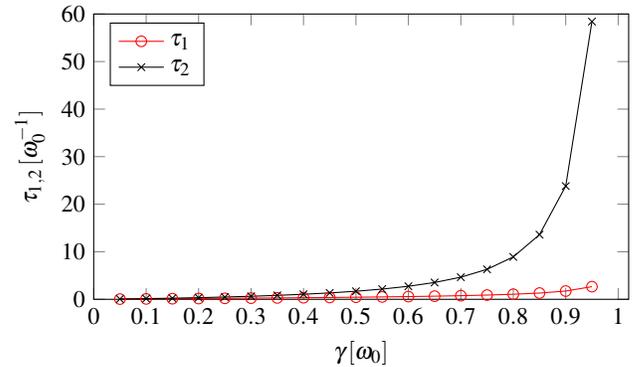
Fig.\ \ref{fig:Fig2} shows the results obtained from \eqref{tau12}. In the case of $\gamma=0.9\omega_0$: the crossing times are $\tau_1=1.73\omega_0^{-1}$ and $\tau_2=23.81\omega_0^{-1}$, from which we get $A(\tau_1)=0.48x_0$ and $A(\tau_2)=1.14\times10^{-9}x_0$. In this case, for $t > \tau_1$, the underdamped solution $x_{ud}(t)$ oscillates with an envelope that is smaller in magnitude than the critically damped solution, until both solutions reach a displacement of the order $10^{-9}x_0$ at $t=\tau_2$. In general, since the magnitude of the underdamped solution's displacement is smaller than the critically damped solution for $\tau_1 < t < \tau_2$, then if $A(\tau_2)$ is less than the experimental resolution, the underdamped displacement will not be measured to be larger than the critically damped displacement at any time after $\tau_1$. Note that $\tau_2$ diverges as $\gamma$ approaches $\omega_0$, and therefore $A(\tau_2)$ can be made arbitrarily small by choosing a large enough damping coefficient. Therefore, even though the critically damped displacement approaches equilibrium faster than the underdamped displacement for $t > \tau_2$, for any given experimental resolution, one can find a damping coefficient $\gamma$ for which the underdamped solution will be measured to reach equilibrium first.
\begin{figure}[h!t!]
\begin{center}
\begin{tikzpicture}
        \begin{axis}[
        width=0.485\textwidth,
        height=0.3\textwidth,
        xmin = 0,
        xmax = 10,
        ymin = -9,
        ymax = 1,
        xtick={0,1,...,9,10},
        ytick={-8,-6,...,-2,0},
        every tick label/.append style={font=\small},
        ylabel near ticks,
        xlabel near ticks,
        xlabel = {\small $t\,[\omega_0^{-1}]$},
        ylabel = {\small $\log_{10} \left(E(t)/E_0\right)$},
        legend entries = {\footnotesize $\gamma=0.85\omega_0$\\\footnotesize $\gamma=0.9\omega_0$\\\footnotesize $\gamma=0.95\omega_0$\\\footnotesize $\gamma=\omega_0$\\}
        ]
        \addplot [domain=0:10,samples=200,thick, dash dot, green](\x,{-0.738*\x+0.557+log10(sin(deg(0.527*\x))^2+cos(deg(0.527*\x-1.016))^2)});
        \addplot [domain=0:10,samples=200,thick, red](\x,{-0.782*\x+0.721+log10(sin(deg(0.436*\x))^2+cos(deg(0.436*\x-1.119))^2)});
        \addplot [domain=0:10,samples=200,thick, blue, densely dotted](\x,{-0.825*\x+1.011+log10(sin(deg(0.312*\x))^2+cos(deg(0.312*\x-1.253))^2)});
        \addplot [domain=0:10,samples=100,thick,dashed, black](\x,{log10((1+\x)^2+\x^2)-0.869*\x});
        \addplot [domain=0:10,samples=100,dashed, black](\x,{-5.5});
        
        \node[] at (axis cs: 1.35,-1.5) {(a)};
        \node[] at (axis cs: 2,-4.9) {\small $E(t) = 10^{-5.5}E_0$};
        \end{axis}  
    \end{tikzpicture}
\end{center}
\begin{center}
\begin{tikzpicture}
        \begin{axis}[
        width=0.485\textwidth,
        height=0.3\textwidth,
        xmin = 0,
        xmax = 50,
        ymin = -40,
        ymax = 3,
        xtick={0,10,...,40,50},
        ytick={-40,-30,...,-10,0},
        every tick label/.append style={font=\small},
        ylabel near ticks,
        xlabel near ticks,
        xlabel = {\small $t\,[\omega_0^{-1}]$},
        ylabel = {\small $\log_{10} \left(E(t)/E_0\right)$},
        legend entries = {\footnotesize $\gamma=0.9\omega_0$\\\footnotesize $\gamma=\omega_0$\\}
        ]
        \addplot [domain=0:50,samples=200,thick, red](\x,{-0.782*\x+0.721+log10(sin(deg(0.436*\x))^2+cos(deg(0.436*\x-1.119))^2)});
        \addplot [domain=0:50,samples=100,thick,dashed, black](\x,{log10((1+\x)^2+\x^2)-0.869*\x});
        
        \node[] at (axis cs: 6,-15) {(b)};
        
        \draw[->,>=stealth] (axis cs: 24.2,-25) -- (axis cs: 24.2,-18);
        \node[] at (axis cs: 24.2,-28) {\small $E(t) \approx 10^{-18}E_0$};
        
        \draw[->,>=stealth] (axis cs: 43,-26) -- (axis cs: 43,-33);
        \node[] at (axis cs: 43,-24.5) {\small $E_c < E_{ud}$};
        
        \end{axis}  
    \end{tikzpicture}
\end{center}
\caption{(a)\,The base 10 logarithm of the energy ratio \eqref{E}, for underdamped oscillators with $\gamma=0.85\omega_0$ (green dash-dotted curve), $\gamma=0.9\omega_0$ (red solid curve), $\gamma=0.95\omega_0$ (blue dotted curve), and for the critically damped oscillator (black dashed curve). (b)\,The energy ratio \eqref{E} for $\gamma=0.9\omega_0$ (red solid curve) and for the critically damped oscillator (black dashed curve) during a longer time period. The energies of the two systems are equal when $t=24.24\omega_0^{-1}$, as indicated by an arrow. After this moment, the energy ratio for the underdamped case oscillates above and below the same ratio for the critically damped oscillator, until $t=43.03\omega_0^{-1}$ (also indicated with an arrow), after which it always remains above it.}
\label{fig:Fig3}
\end{figure}
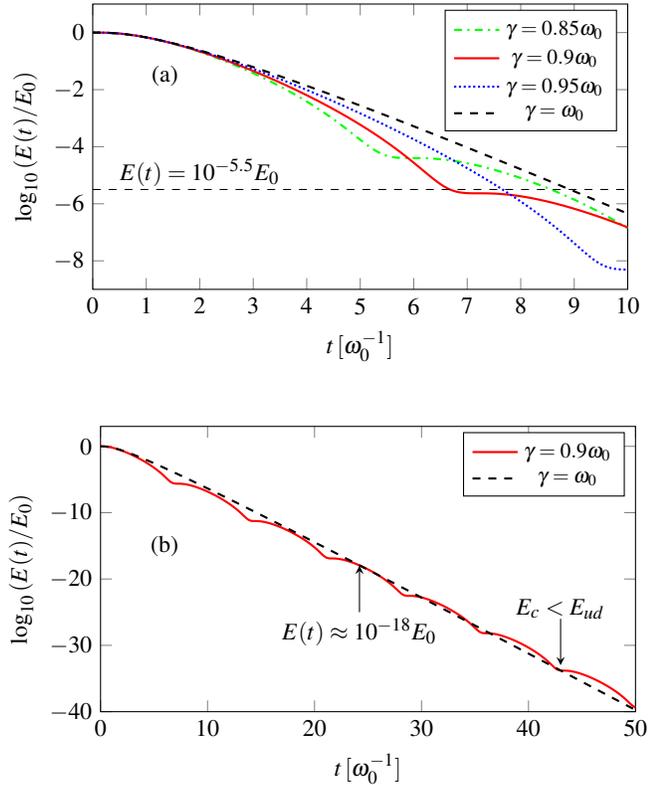

\subsection{Energy analysis}
\label{subsection:energy} 

We now turn our attention to the energy of the system. In Fig.\,\ref{fig:Fig3}\,(a)\,we show the base 10 logarithm of the ratio \eqref{E}, for four values of $\gamma=\lbrace0.85\omega_0, 0.9\omega_0, 0.95\omega_0, \omega_0\rbrace$. The energy ratios for various underdamped cases are only slightly larger than the corresponding ratio for the critically damped oscillator from $t=0$ until $t=\lbrace{1.10\omega_0^{-1},1.06\omega_0^{-1},1.03\omega_0^{-1}\rbrace}$, for $\gamma=\lbrace{0.85\omega_0, 0.9\omega_0, 0.95\omega_0\rbrace}$ respectively.  After these moments, the underdamped curves remain lower than the critically damped one for the rest of the time period of Fig.\ \ref{fig:Fig3}(a). In Fig.\ \ref{fig:Fig3}(b)\,we show the same behavior, this time for the case when $\gamma=0.9\omega_0$, and during a longer time period. We note that the energy ratios of the two systems are equal when $t=24.24\omega_0^{-1}$. This energy ratio equality occurs slightly later in time than the equality of the underdamped oscillator envelope and the critically damped oscillator displacement, ($t=\tau_2=23.81\omega_0^{-1}$ as shown in Sec. \ref{subsection:displacement}). We expect that these times do not coincide due to the differences between the envelope and the full underdamped solution, i.e., the underdamped displacement exceeds the critical displacement in magnitude later than the envelope. After $t=24.24\omega_0^{-1}$, the energy ratio of the underdamped oscillator periodically dips below that of the critically damped oscillator.  Finally, when $t>43.03\omega_0^{-1}$, which corresponds to  $E(t)<10^{-33}E_0$, the energy ratio of the underdamped oscillator remains permanently above the energy ratio for the critically damped oscillator, in accordance with infinite time limit \eqref{xc_xud}. Other values of the underdamped damping coefficient show the same qualitative behavior. To determine an optimal damping coefficient $\gamma$ for which the system's energy will drop to a desired energy level the soonest, we first give an example in Fig.\,\ref{fig:Fig3}(a), where we draw a horizontal line corresponding to $E(t)/E_0=10^{-5.5}$. It is immediately clear that the underdamped oscillator with $\gamma=0.9\omega_0$ reaches the line corresponding to $E(t)/E_0=10^{-5.5}$ first, before the other underdamped oscillators.

\subsection{Existence of a unique optimal damping coefficient}
\label{subsection:existence}

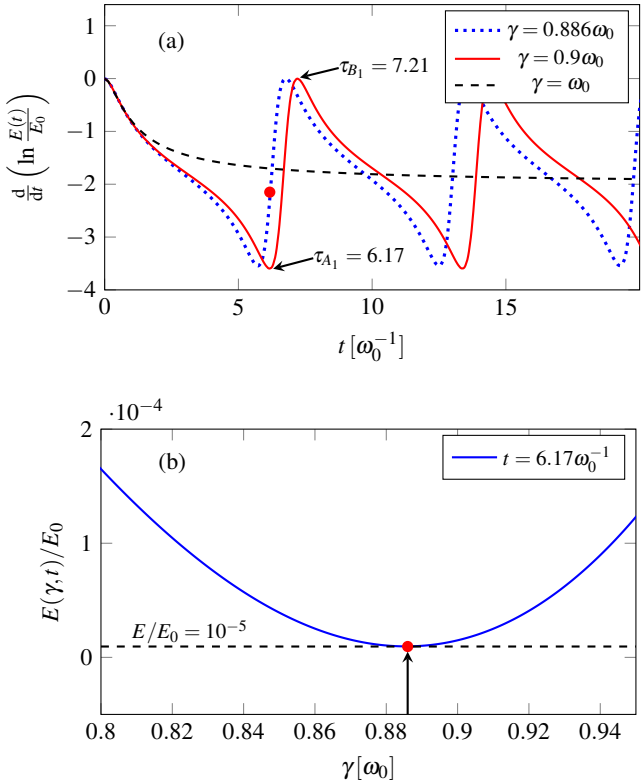
\begin{figure}
\begin{center}
\hskip -4mm
\begin{tikzpicture}
        \begin{axis}[
        width=0.485\textwidth,
        height=0.3\textwidth,
        xmin = 0,
        xmax = 20,
        ymin = -4,
        ymax = 1.4,
        xtick={0,5,...,15},
        ytick={-4,-3,...,0,1},
        every tick label/.append style={font=\small},
        ylabel near ticks,
        xlabel near ticks,
        xlabel = {\small $t\,[\omega_0^{-1}]$},
        ylabel = {\small $\frac{\textrm{d}}{\textrm{d}t}\left(\ln\frac{E(t)}{E_0}\right)$},
        legend entries = {\footnotesize $\gamma=0.886\omega_0$\\\footnotesize $\gamma=0.9\omega_0$\\\footnotesize $\gamma=\omega_0$\\}
        ]
        \addplot [smooth,domain=0:26,samples=200,very thick, blue, dotted](\x,{-4*0.886*(2.157*sin(deg(0.464*\x)))^2/((2.157*sin(deg(0.464*\x)))^2+(cos(deg(0.464*\x))+1.911*sin(deg(0.464*\x)))^2)});
        \addplot [smooth,domain=0:26,samples=200,thick, red](\x,{-1.8+0.872*(sin(deg(0.436*\x))*cos(deg(0.436*\x))-cos(deg(0.436*\x-1.119))*sin(deg(0.436*\x-1.119)))/(sin(deg(0.436*\x))^2+cos(deg(0.436*\x-1.119))^2)});
        \addplot [domain=0:26,samples=200,thick, dashed](\x,{-2+2*(1+2*\x)/((1+\x)^2+\x^2)});

        \node[] at (axis cs: 2.5,0.7) {(a)};
        \node[fill, red, circle,inner sep=1.5pt,label={}] at (6.1726,-2.15){};
                
        \draw[->,>=stealth, thick] (axis cs: 7.8,-3.4) -- (axis cs: 6.3,-3.6);
        \node[] at (axis cs: 9.5,-3.4) {\footnotesize $\tau_{A_1}=6.17$};
        \draw[->,>=stealth, thick] (axis cs: 8.75,0.2) -- (axis cs: 7.3,0.02);
        \node[] at (axis cs: 10.45,0.2) {\footnotesize $\tau_{B_1}=7.21$};

        \end{axis}  
    \end{tikzpicture}
\end{center}
\begin{center}
\begin{tikzpicture}
        \begin{axis}[
        width=0.485\textwidth,
        height=0.3\textwidth,
        xmin = 0.8,
        xmax = 0.95,
        ymin = -0.00005,
        ymax = 0.0002,
        xtick={0.8,0.82,...,0.94},
        ytick={0,0.0001,0.0002},
        every tick label/.append style={font=\small},
        ylabel near ticks,
        xlabel near ticks,
        xlabel = {\small $\gamma\,[\omega_0]$},
        ylabel = {\small $E(\gamma,t)/E_0$},
        legend entries = {\footnotesize $t=6.17\omega_0^{-1}$\\}
        ]
        \addplot [smooth, domain=0.7:0.999,samples=200, thick, blue](\x,{exp(-2*\x*6.1726)*((cos(deg(sqrt(1-\x^2)*6.1726)))^2+\x*sin(deg(2*sqrt(1-\x^2)*6.1726))/sqrt(1-\x^2)+(1+\x^2)*(sin(deg(sqrt(1-\x^2)*6.1726)))^2/(1-\x^2))});
        \node[fill, red, circle,inner sep=1.5pt,label={}] at (0.886,0.0000095){};
        \addplot [domain=0.8:0.884,samples=50,thick, dashed](\x,{0.0000095});
        \addplot [domain=0.8875:0.95,samples=50,thick, dashed](\x,{0.0000095});
        \node[] at (axis cs: 0.825,0.0000225) {\footnotesize $E/E_0=10^{-5}$};
        \draw[->,>=stealth, thick] (axis cs: 0.886,-0.00005) -- (axis cs: 0.886,0.000005);
       \node[] at (axis cs: 0.82,0.00017) {(b)};
        \end{axis}  
    \end{tikzpicture}
\end{center}
\caption{(a) The energy loss rates \eqref{lossln} of the critically damped oscillator (black dashed curve) and underdamped oscillators with $\gamma=0.886\omega_0$ (blue dotted curve) and $\gamma=0.9\omega_0$ (red solid curve). The first equilibrium crossing time $\tau_{A_1}$ and first time the system reaches a turning point $\tau_{B_1}$ are indicated with arrows. For $\gamma=0.9\omega_0$ we have $\tau_{A_1}=6.17\omega_0^{-1}$, $\tau_{B_1}=7.21\omega_0^{-1}$. The red circle indicates the value of the energy loss rate for $\gamma=0.886\omega_0$ at $t=6.17\omega_0^{-1}$. (b) Energy ratio \eqref{E} as a function of $\gamma$ at $t=6.17\omega_0^{-1}$ (blue solid curve). The energy ratio has a global minimum for $\gamma=0.886\omega_0$ (the arrow points to the minimum indicated by the red circle). The dashed horizontal line indicates the energy level reached by $\gamma=0.886\omega_0$ at $t=6.17\omega_0^{-1}$. See text for details.}
\label{fig:Fig4}
\end{figure}

Using a particular example, we now show that it is possible to determine a unique underdamped $\gamma$ for which the oscillator reaches some energy level faster than for any other $\gamma$. The underdamped oscillator, described by \eqref{uds1}, is in the equilibrium position at moments given by
\begin{equation}
\tau_{A_n}=\frac{1}{\omega}\left[(2n-1)\frac{\pi}{2}+\arctan\left(\frac{\gamma}{\omega}\right)\right]\,,
\label{tauA}
\end{equation}
and is located at a turning point at times
\begin{equation}
\tau_{B_n}=\frac{n\pi}{\omega}\,,
\label{tauB}
\end{equation}
where $n\in\mathbb{N}$ is the index that counts the equilibrium crossings and turning points respectively. In Fig.\,\ref{fig:Fig4}(a) we show the energy loss rates \eqref{lossln} of the critically damped oscillator and underdamped oscillators for $\gamma=\lbrace 0.886\omega_0, 0.9\omega_0\rbrace$. For the case $\gamma=0.9\omega_0$ the maximal magnitude of the energy loss rate is $4\gamma=3.6\omega_0$. This means, according to \eqref{lossln}, $\beta_{ud}=1$ at time $\tau_{A_1}=6.17\omega_0^{-1}$. At this moment, the system first passes through the equilibrium position. By contrast, the energy loss rate is zero when $\beta_{ud}=0$ at $\tau_{B_1}=7.21\omega_0^{-1}$, when the system arrives for the first time at a turning point. It is clear that $\gamma=0.886\omega_0$ (or any other $\gamma<\omega_0$), yields qualitatively the same behavior.

In Fig.\,\ref{fig:Fig4}(b) we show the energy ratio \eqref{E} as a function of $\gamma$ at $t=6.17\omega_0^{-1}$. At this moment, the oscillator with $\gamma=0.9\omega_0$ reaches the equilibrium position for the first time, and we can see that the energy ratio has a global minimum for a slightly smaller damping coefficient, i.e. for $\gamma=0.886\omega_0$. The value of this energy minimum is $10^{-5}E_0$. Let's assume that this is the energy resolution of the experiment; we denote this threshold energy as $E_{th}$. We note here that, for any $\gamma$, the energy is a monotonically decreasing function of time. Thus, $E(\gamma,t)>E_{th}$ for $t<6.17\omega_0^{-1}$, and the oscillator with $\gamma=0.886\omega_0$ reaches $E_{th}$ at $t=6.17\omega_0^{-1}$, faster than oscillators with any other $\gamma$. We can conclude that $\gamma=0.886\omega_0$ is the optimal damping coefficient for experiment with $E_{th}=10^{-5}E_0$.

This example shows us that if, at some instant, the energy as a function of $\gamma$ has a global minimum at some $\gamma_{min}$, the oscillator with $\gamma_{min}$ reaches that energy minimum faster than all other oscillators, i.e., $\gamma_{min}$ is the optimal damping coefficient for that energy threshold.
In the appendix \ref{globalmin} we present the behavior of the energy ratio \eqref{E} and its global minimum for a wider range of $\gamma$, and at different instants over a larger time span. It is clear that for any energy level $E_{th}$ the unique optimal damping coefficient can be found. 

We note here that the oscillator with optimal damping coefficient reaches $E_{th}$ when its displacement is between the equilibrium position and the first turning point (designated by the red circle in Fig.\,\ref{fig:Fig4}(a)). At the same instant, the oscillator with slightly larger damping coefficient, i.e. with $\gamma=0.9\omega_0$, arrives for the first time at the equilibrium position and has slightly more energy than the optimally damped oscillator (see Fig.\,\ref{fig:Fig4}(b)). In appendix \ref{globalmin} we show that this difference---between the optimal damping coefficient (for which the energy ratio \eqref{E} has a minimum at some instant) and the damping coefficient which causes the system to first reach equilibrium at that same instant---decreases as we move to lower values of $E_{th}$. Therefore, as we show below, if we use \eqref{tauA} to determine $\gamma$ for which the system first comes to equilibrium precisely when its energy is equal to $E_{th}$, we have obtained an excellent first approximation of the optimal damping coefficient and a good starting point for its precise numerical determination.

\subsection{Determination of the optimal damping coefficient}
\label{subsection:optimum}

We define the system to be effectively at equilibrium when its energy equals $10^{-\delta}E_0$, with some $\delta>0$ and we find the optimal underdamped coefficient $\gamma_{opt}$ in two steps. 

First we determine $\gamma$ for which the system reaches this energy level when passing through equilibrium for the first time. We will show that this value of $\gamma$ is an excellent first approximation for the optimal damping coefficient. Therefore, we will refer to it as the \underline{first step optimal damping} denoted as $\gamma_1$. In the second step, we use $\gamma_1$ to find a small subset of underdamped coefficients for which the system reaches energy $10^{-\delta}E_0$ faster than the system with $\gamma_1$, and we then determine the optimal damping coefficient $\gamma_{opt}$ from that subset.   

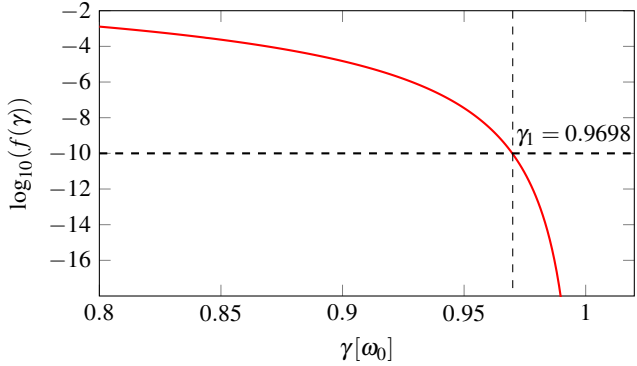
\begin{figure}[h!t!]
\begin{center}
\begin{tikzpicture}
        \begin{axis}[
        width=0.485\textwidth,
        height=0.3\textwidth,
        xmin = 0.8,
        xmax = 1.02,
        ymin = -18,
        ymax = -2,
        xtick={0.8,0.85,0.9,0.95,1},
        ytick={-16,-14,...,-4,-2},
        every tick label/.append style={font=\small},
        ylabel near ticks,
        xlabel near ticks,
        xlabel = {\small $\gamma\,[\omega_0]$},
        ylabel = {\small $\log_{10}(f(\gamma))$}
        ]
        \addplot [domain=0.8:0.99,samples=200,thick, red](\x,{log10(exp(
        (-2*\x)*(1.57+rad(atan(\x/(sqrt(1-\x*\x)))))/(sqrt(1-\x*\x))))});
        \addplot [domain=0.8:1.02,samples=50,thick, dashed](\x,{-10});
        
        \node[] at (axis cs: 0.995,-9) {\small $\gamma_1=0.9698$};
        \draw[dashed] (axis cs: 0.97,-2) -- (axis cs: 0.97,-18);
        
        \end{axis}  
    \end{tikzpicture}
\end{center}
\caption{Graphical determination of the first step optimal damping $\gamma_1$ for $\delta=10$ using condition \eqref{condition3}. For the determined $\gamma_1 = 0.9698\omega_0$ the energy of the underdamped oscillator permanently drops below the preset threshold energy when arriving at the equilibrium position for the first time.}
\label{fig:Fig6}
\end{figure}

The first step optimal damping coefficient $\gamma_1$ is determined by the condition $E(\tau_{A1}) = 10^ {-\delta}E_0$. Since at $\tau_{A_1}$ the total energy is equal to the kinetic energy, we can write the chosen condition as
\begin{equation}
\frac{E_{ud}(\tau_{A_1})}{E_0}=\frac{\dot x_{ud}(\tau_{A_1})^2}{\omega_0^2x_0^2}=10^{-\delta}\,.
\label{condition1}
\end{equation}
Next, using \eqref{tauA} with $n=1$ and the time derivative of \eqref{uds1}, we get
\begin{equation}
\exp\left[-2\frac{\gamma}{\sqrt{\omega_0^2-\gamma^2}}\left(\frac{\pi}{2}+\arctan\frac{\gamma}{\sqrt{\omega_0^2-\gamma^2}}\right)\right]=10^{-\delta}.
\label{condition2}
\end{equation}
The left-hand side of \eqref{condition2} is a function of $\gamma$ only, we denote it as $f(\gamma)$. We can then graphically determine $\gamma_1$, for any given $\delta>0$, from the condition
\begin{equation}
\log_{10}(f(\gamma))=-\delta\,.
\label{condition3}
\end{equation}  
In Fig.\,\ref{fig:Fig6} we plot $\log_{10}f(\gamma)$ and the horizontal line corresponding to $\delta=10$.  Their intersections leads to $\gamma_1 = 0.9698\omega_0$. We give the results for other choices of $\delta$ in Table I.

\begingroup
\setlength{\tabcolsep}{5.2pt} 
\renewcommand{\arraystretch}{1.5} 
\begin{table}[h!t!]
\begin{tabular}{|c|c|c|c|c|} 
 \hline
 $E(t)/E_0$ & $\gamma_1 \left[ \omega_0 \right]$ & $\tau_{A_1} [ \omega_0^{-1}]$ & $\tau_{c}[\omega_0^{-1}]$ & $\left(\tau_{c} - \tau_{A_1} \right)/\tau_{c} \left[ \% \right]$ \\ 
 \thickhline
 $10^{-4}$ & \ $0.8688$ \ & $5.30$ & $6.96$ & $23.85$ \\ 
 \hline
 $10^{-6}$ & \ $0.9286$ \ & $7.44$ & $9.56$ & $22.17$ \\ 
 \hline
 $10^{-8}$ & \ $0.9555$ \ & $9.63$ & $12.10$ & $20.35$ \\ 
 \hline
 $10^{-10}$ & \ $0.9698$ \ & $11.87$ & $14.57$ & $18.53$ \\ 
 \hline
 $10^{-12}$ & \ $0.9782$ \ & $14.12$ & $17.03$ & $17.09$ \\ 
 \hline
 $10^{-14}$ & \ $0.9835$ \ & $16.36$ & $19.46$ & $15.98$ \\ 
 \hline
 $10^{-16}$ & \ $0.9872$ \ & $18.69$ & $21.88$ & $14.54$ \\ 
 \hline
 $10^{-18}$ & \ $0.9897$ \ & $20.94$ & $24.28$ & $13.76$ \\ 
 \hline
\end{tabular}
\caption{The first column contains the ratio $E(t)/E_0$ which defines the effective equilibrium state. The second column gives the first step optimal underdamped coefficient $\gamma_1$ obtained from \eqref{condition3}. The third column contains the time $\tau_{A_1}$ at which the underdamped oscillator (with coefficient $\gamma_1$) has reached the effective equilibrium state. The fourth column contains the time at which the critically damped oscillator reaches that same level of energy, and the fifth column contains the relative time advantage of the underdamped oscillator compared to the critically damped oscillator.}
\end{table}
\endgroup

\begin{figure}[h!t!]
\begin{center}
\begin{tikzpicture}
        \begin{axis}[
        width=0.485\textwidth,
        height=0.3\textwidth,
        xmin = 0.87,
        xmax = 0.96,
        ymin = -0.0000025,
        ymax = 0.00001,
        xtick={0.85,0.87,...,0.97},
        ytick={0,0.00001,0.00002},
        every tick label/.append style={font=\small},
        ylabel near ticks,
        xlabel near ticks,
        xlabel = {\small $\gamma\,[\omega_0]$},
        ylabel = {\small $E(\gamma,t)/E_0$},
        legend entries = {\footnotesize $t=7.44\omega_0^{-1}$\\\footnotesize $t=7.20\omega_0^{-1}$\\}
        ]
        \addplot [smooth, domain=0.7:0.999,samples=200,thick, red](\x,{exp(-2*\x*7.44)*((cos(deg(sqrt(1-\x^2)*7.44)))^2+\x*sin(deg(2*sqrt(1-\x^2)*7.44))/sqrt(1-\x^2)+(1+\x^2)*(sin(deg(sqrt(1-\x^2)*7.44)))^2/(1-\x^2))});
        \addplot [dashed, domain=0.7:0.999,samples=200,thick, blue](\x,{exp(-2*\x*7.2035)*((cos(deg(sqrt(1-\x^2)*7.2035)))^2+\x*sin(deg(2*sqrt(1-\x^2)*7.2035))/sqrt(1-\x^2)+(1+\x^2)*(sin(deg(sqrt(1-\x^2)*7.2035)))^2/(1-\x^2))});
        \addplot [domain=0.8:1,samples=50,thick, dashed](\x,{0.000001});
        \node[] at (axis cs: 0.885,0.00000175) {\footnotesize $E/E_0=10^{-6}$};
        \draw[->,>=stealth, thick] (axis cs: 0.9286,-0.0000025) -- (axis cs: 0.9286,0.000001);
         \draw[->,>=stealth, thick, dashed] (axis cs: 0.915,-0.0000025) -- (axis cs: 0.915,0.000001);
        \node[] at (axis cs: 0.82,0.00017) {(b)};
        \end{axis}  
    \end{tikzpicture}
\end{center}
\caption{Energy ratio \eqref{E} as a function of $\gamma$ at $t=7.44\omega_0^{-1}$ (red solid curve) and $t=7.20\omega_0^{-1}$ (blue dashed curve). The energy level reached by $\gamma=0.9286\omega_0$ at $t=7.44\omega_0^{-1}$ is indicated by the dashed horizontal line, and the point corresponding to $\gamma=0.9286\omega_0$ on the red solid curve is indicated by the solid arrow. The part of the red curve, under the horizontal line, corresponds to the interval $\gamma\in(0.9104\omega_0,0.9286\omega_0)$ for which (at this instant) the energy is lower than the energy for $\gamma=0.9286\omega_0$. For $\gamma=0.9145\omega_0$ the energy at $t=7.20\omega_0^{-1}$ has minimum indicated by the dashed arrow. See text for details.}
\label{fig:Fig7}
\end{figure}
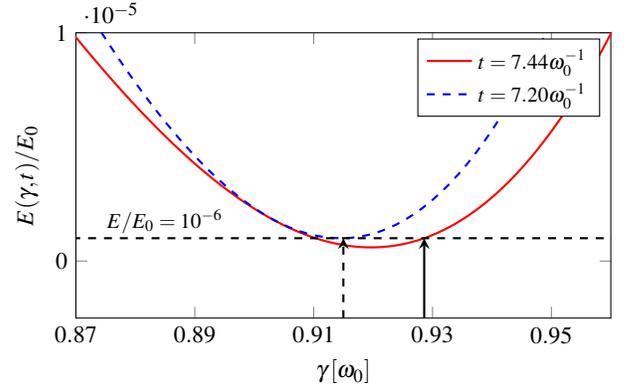

We now proceed to the second step and find the optimal damping coefficient. We focus on the energy resolution $10^{-6}E_0$; for any other choice the procedure is the same. From Table I, we see that the oscillator with $\gamma_1=0.9286\omega_0$ reaches this resolution when first at equilibrium position, i.e., at $t=7.44\omega_0^{-1}$. In Fig.\,\ref{fig:Fig7} we show the energy ratio \eqref{E} as a function of $\gamma$ at this instant. From this graph we determine the subset of underdamped coefficients $\gamma\in(0.9104\omega_0,0.9286\omega_0)$ with energy smaller than $10^{-6}E_0$ at this instant; this is the subset of underdamped coefficients that drop below the experimental energy resolution faster than $\gamma_1$. Since this subset is only $\Delta\gamma=0.0182\omega_0$ wide, we determined the optimal damping coefficient simply by stepping through the values in this subset and numerically finding the value of $\gamma$ which causes the system's energy ratio \eqref{E} to intersect the level $10^{-6}$ in the shortest time. We find that $\gamma_{opt}=0.9145\omega_0$ reaches the energy resolution at $t=7.20\omega_0^{-1}$, faster than any other $\gamma$. To confirm this, in Fig.\,\ref{fig:Fig7} we plot the energy ratio \eqref{E} as a function of $\gamma$ at $t=7.20\omega_0^{-1}$, and we see that the minimum for $\gamma_{opt}=0.9145\omega_0$ touches the energy resolution level. 

In this example, $\gamma_{opt}$ is $1.52\%$ smaller than $\gamma_1$ (this difference decreases as the energy threshold is lowered, see appendix \ref{globalmin}) and the system takes $3.26\%$ less time than the system with $\gamma_1$ to reach the energy resolution level. Depending on the experimental conditions, this can be significant, but for many applications it suffices to take $\gamma_1$ as optimal. For example, in the next section, where we experimentally test our findings, the difference between $\gamma_{opt}$ and $\gamma_1$ is undetectable since it results in a change that is within the experimental resolution.

For all other types of initial conditions, the system's energy behavior is qualitatively the same as for the initial conditions $x_0>0$ and $v_0=0$ (see supplementary material \cite{SupplMat}, sections SII-SIV), with one exception. For the special case in which the initial displacement and velocity are of opposite signs and satisfy $|v_0|>|x_0\omega_0|$, one can choose 
\begin{equation}
\gamma=\frac{\left(\frac{|v_0|}{|x_0|}\right)^2+\omega_0^2}{2\frac{|v_0|}{|x_0|}} \, 
\label{gammatilda}
\end{equation}
in the overdamped regime, for which the factor multiplying the slowly decaying exponential in \eqref{ods} vanishes, leaving only the quickly decaying exponential $e^{(-\gamma-\alpha)t}$. For this particular initial condition, the overdamped solution with the damping coefficient \eqref{gammatilda} returns to equilibrium sooner than any other solution (details are given in supplementary material \cite{SupplMat}, section SIV, subsection C). 

For the interested reader, in the supplementary material \cite{SupplMat}, sections SV and SVI, we examine the relationship between the energy loss rates and the ratio of the critically damped to underdamped energy and we show the results of the optimisation with respect to $\tau_{A_2}$, i.e., allowing the system to pass the equilibrium position once, overshoot it, and reach the experimental energy resolution when coming into the equilibrium position the second time. It is expected that the advantage of the underdamped oscillator over the critically damped oscillator is smaller for the results obtained this way, since the energy of the underdamped oscillator gets closer to the energy of the critically damped oscillator after each cycle, until it eventually surpasses it (see Fig.\,3(a) and (b)).

\section{Experimental verification}

We tested our model using a series RLC circuit. Other systems are also suitable for this purpose, such as a physical pendulum with an eddy-current damping system that allows the damping conditions to be controlled with great precision \cite{Manuel}.

The differential equation describing the circuit is
\begin{equation}
\frac{\textrm{d}^2I(t)}{\textrm{d}t^2}+\frac{R}{L}\frac{\textrm{d}I(t)}{\textrm{d}t} + \frac{1}{LC}I(t) = 0 \,.
\end{equation}
This equation has the form of a damped harmonic oscillator \eqref{DHOeq}, with
\begin{equation}
\omega_0^2 = \frac{1}{LC} \hskip 5mm \textrm{and} \hskip 5mm  2\gamma = \frac{R}{L} \,.
\end{equation}
Since in this work we always expressed $\gamma$ in units of $\omega_0$, it is more convenient to write
\begin{eqnarray}
\gamma = \frac{R}{2L}(\omega_0\sqrt{LC}) = \frac{R}{2}\sqrt{\frac{C}{L}}\omega_0 = \zeta \omega_0\,.
\end{eqnarray}
We note that the current in the circuit $I(t)$ exhibits damped oscillations, but so does the voltage across the resistor $RI(t)$, which can be measured with an oscilloscope.

The circuit was constructed using a decade resistor with a range from $0\,\Omega$ to $1\,\textrm{M}\Omega$ with a manufacturer stated tolerance of 1\%, an inductor with inductance of $(14.80 \pm 0.03 )\,\textrm{mH}$ and a variable capacitor with a range from 0\,F to 10\,$\mu$F. The voltage was provided by a square wave from an Agilent 33500B waveform generator with low frequency distortion ($\pm$ 1\,$\mu$Hz) and high voltage retention (1\% of setting) \cite{Agilent}. The voltage across the resistor was measured using a UNI-T UTD2102CEX 100\,MHz oscilloscope. The oscilloscope's measurement uncertainty depends on the readout scale and was set to 1/4 of the smallest readout unit for both time and voltage. 

\begin{figure}[h!]
\centering
\begin{tikzpicture}
\node[above right] (img) at (0,0) {\includegraphics[width=0.45\textwidth]{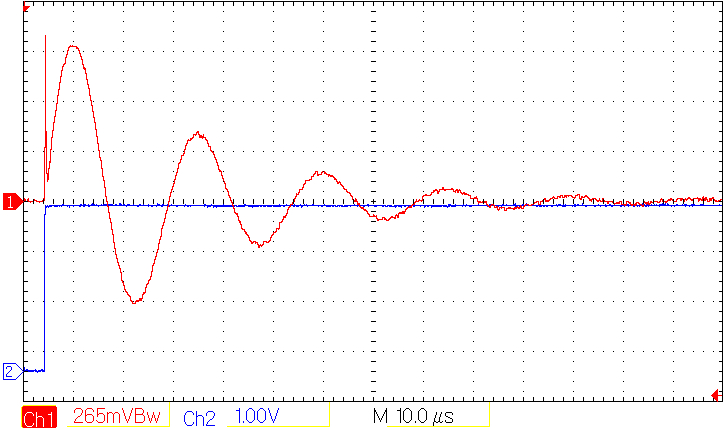}};
\node at (190pt,120pt) {\large $R=800\,\Omega$};
\end{tikzpicture}
\caption{An example of the underdamped oscillating voltage measured across the resistor with circuit parameters $R=800\,\Omega$, $L=14.8\,\textrm{mH}$ and $C=1\,\textrm{nF}$. Voltage is plotted on the vertical axis, and time is plotted on the horizontal axis.}
\label{fig:frekvencija}
\end{figure}

Initially, the wave source frequency was set to 200\,Hz with a voltage step size of 3.3\,V. The capacitance was set to 1\,nF and the resistance to 800\,$\Omega$. We  verified that the circuit had no significant stray resistances and inductances with the PeakTech 2005 and Fluke 115 multimeters. We observed underdamped oscillations of the voltage across the resistor, shown in Fig.\,\ref{fig:frekvencija}.  The blue line is the voltage from the switching voltage source (unit voltage step of 1\,V) and the red line is the voltage measured across the resistor (unit voltage step of 265\,mV). Both are measured on the same time axis with a unit measuring step of $10\,\mu\textrm{s}$.

Using the oscilloscope's measuring tools, we took five measurements of the time interval between two successive maxima or minima of the oscillations and took the inverse to obtain the frequency which was found to be $(40.0 \pm 0.2)\,\textrm{kHz}$. Using this value, we determined the circuit's capacitance and obtained $(1.06 \pm 0.01)\,\textrm{nF}$. This is slightly different from the value set on the variable capacitor due to unavoidable stray capacitances in the circuit. Using these values, we can now find the dimensionless damping factor $\zeta$ of the circuit
\begin{equation}
\zeta = \frac{R}{2}\sqrt{\frac{C}{L}} = (0.107 \pm 0.001)\,,
\end{equation}
which confirms that the oscillations take place well in the underdamped regime. The value of the resistance that corresponds to critically damped oscillations is:
\begin{equation}
R_{critical} = 2\sqrt{\frac{L}{C}} = (7.47 \pm 0.04)\,\textrm{k}\Omega\,.
\end{equation} 

At this point we note that the calculated uncertainty of $R_{critical}$ is purely a consequence of measurement uncertainties and does not reflect the experimental resolution. Upon examination, we saw that a change in the resistance of the setup which was smaller than 250\,$\Omega$ in either direction resulted in a voltage change that was within the experimental resolution, which puts the resolution at the level of $\approx 7\%$.

\begin{figure}[h!]
\centering
\begin{tikzpicture}
\node[above right] (img) at (0,0) {\includegraphics[width=0.45\textwidth]{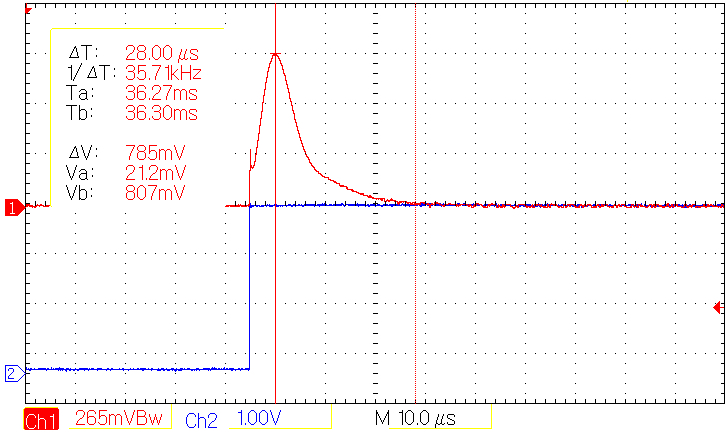}};
\node at (190pt,120pt) {\large $R=7470\,\Omega$};
\end{tikzpicture}
\caption{The oscilloscope readout in the case of the critically damped circuit with $R=7.47\,\textrm{k}\Omega$ (red curve). The time interval required for the voltage to drop from the maximal value to the resolution of the oscilloscope was  $(28.0 \pm 0.4)$\,$\mu$s, which is the interval between the vertical red lines. The blue line is the switching voltage signal.}
\label{fig:kriticni}
\end{figure}

After setting the value of the resistance to the critical value, we performed the measurement of the voltage across the resistor again. No oscillations were visible this time, as can be seen in Fig.\ref{fig:kriticni}, which is the expected behavior of a critically damped oscillator.

We measured the time interval between the moment when the signal was at its maximum and the moment when it became indistinguishable from the oscilloscope's measurement resolution, which was verified by the measuring tool. This resolution is taken to be 1/4 of the smallest measuring increment, which in this case corresponds to $\approx 13\,$mV. Since the maximum voltage was $\approx 800\,$mV, the resolution is approximately 1.625\% of the maximal voltage. The measured time interval between these two moments was $(28.0\pm 0.4)$\,$\mu$s $=(7.1 \pm 0.1)$\,$\omega_0^{-1}$. The uncertainty in the measurement is a consequence of the time resolution of the measuring tool, which has a single step of $\Delta t = 0.2$\,$\mu$s.

Since we measured the time interval starting from the moment the oscillator was at its maximum displacement. This, in effect, sets the initial condition to that of section \ref{section:x0}, so we can compare our measurements to the results obtained there. Looking at Fig.\ref{fig:Fig3}(a), we note that at $t=(7.1 \pm 0.1)\,\omega_0^{-1}$, the energy of the critically damped oscillator is in the interval $[6.6,9.4] \cdot 10^{-5}\,E_0$. The underdamped $\gamma_{1}=\zeta_{1} \omega_0$ for the center of this interval is obtained from \eqref{condition2}, which translates to a value of resistance of 
\begin{equation}
R_{1} = 2\zeta_{1}\sqrt{\frac{L}{C}} = (6.52 \pm 0.03)\,\textrm{k}\Omega\,.
\label{Roptimal}
\end{equation}
This value was rounded to 6.5\,k$\Omega$ due to the fact that our experimental sensitivity to resistance is of the order of 250 $\Omega$. Note that since the experimental sensitivity to a change in resistance is  $\approx 7\%$, setting the resistance to 6.5\,k$\Omega$ in fact covers a range of $\zeta$ which will include both $\zeta_{1}$, as well as $\zeta_{opt}$, which differs from it at the order of a few percent.

\begin{figure}[h!]
\centering
\begin{tikzpicture}
\node[above right] (img) at (0,0) {\includegraphics[width=0.45\textwidth]{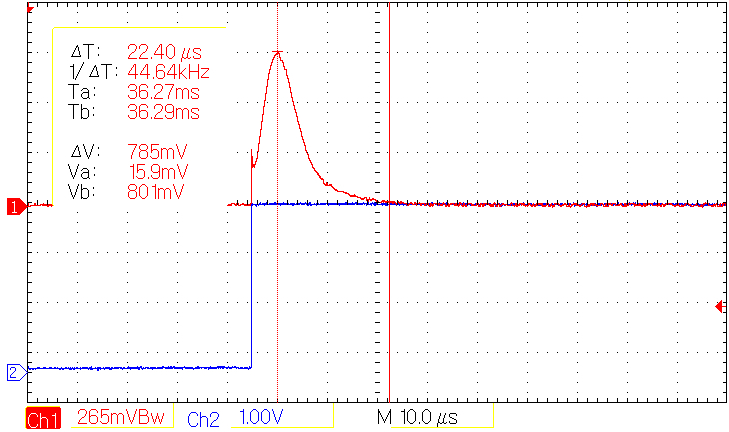}};
\node at (190pt,120pt) {\large $R=6500\,\Omega$};
\end{tikzpicture}
\caption{The oscilloscope readout in the case of the underdamped circuit with $R=6.5\,\textrm{k}\Omega$ (red curve). The time interval required for the voltage to drop from the maximal value to the resolution of the oscilloscope was measured to be $(22.4 \pm 0.4)$\,$\mu$s.}
\label{fig:potkr}
\end{figure}

We first set the value of the resistance to the predicted value of the optimal resistance of 6.5\,k$\Omega$. In Fig.\,\ref{fig:potkr} we show the readout from the oscilloscope for this resistance value. The time required for the voltage to drop from its maximum value to the resolution of the oscilloscope was $(22.4 \pm 0.4)$\,$\mu$s. Compared to the critically damped system, this is a decrease in time of $(20 \pm 3)\%$, in accordance with the findings given in Table I.

Next, we systematically changed the resistance in steps of 250\,$\Omega$, from $4\,\textrm{k}\Omega$ to $8\,\textrm{k}\Omega$. We always ensured that the maximum voltage of the signal was set to 800\,mV. We confirmed that, within the experimental resolution, the voltage dropped mostly quickly to zero when the resistance was set to 6.5\,k$\Omega$, as predicted by \eqref{Roptimal}. 

Decreasing the value of the resistance below the optimal value, we slowly start to notice the overshoot of the voltage signal under the reference voltage (see supplementary material \cite{SupplMat}, section SVII). For resistances above the critical value, the system becomes overdamped and the time it takes for the signal to drop to the reference level is longer than the critical case.

\section{Conclusion}

The results presented here show that, contrary to popular belief, critical damping is never the best choice if we want the damped oscillator to reach an equilibrium state as soon as possible. For most initial conditions, a unique damping coefficient in the underdamped regime can be determined for which the oscillator reaches the equilibrium state the most quickly. An exception to this occurs for a specific choice of initial conditions, when an overdamped oscillator is the optimal choice.

These findings can be significant for understanding all systems described by the damped harmonic oscillator model, such as shock absorbers in cars \cite{UniPhys}, RLC and other oscillating circuits, measuring devices \cite{Rohit}, etc. Recently, the damped oscillator model was used to characterize the convergence of machine learning algorithms, and a connection was established between the differential equations associated with the algorithms and the damped oscillator model\cite{ML}. Although the algorithms converge to solutions with some finite precision, critical damping has previously been considered desirable in terms of convergence speed \cite{ML}. We envision that our results may be significant in that context as well.

\section{Acknowledgments}

This work was supported by the QuantiXLie Center of Excellence, a project co-financed by the Croatian Government and European Union through the European Regional Development Fund, the Competitiveness and Cohesion Operational Programme (Grant No. KK.01.1.1.01.0004).\\

\section{Author contributions} 

K.L. and N.P. share first authorship. K.L. came up with the idea for the paper and did the work on the theoretical side. N.P. did the work on the experimental side. K.L. and N.P. wrote the paper. D.J. participated in the discussions about theoretical issues and in writing of some parts of the paper.\\

The authors have no conflicts to disclose.

\begin{appendix} 

\section{Lambert function}
\label{AppendixLambert}

In this section we discuss in more detail the solutions of equation \eqref{eqtau}. The solutions to this equation, $\tau_1$ and $\tau_2$, are moments in time for which the envelope $A(t)$ of the underdamped oscillator equals the displacement $x_c(t)$ of the oscillator in the critical regime. First, we note that \eqref{eqtau} can be rewritten as 
\begin{equation}
    \left( \frac{\omega}{\omega_0} + \omega \tau\right) e^{-\left(\omega_0- \gamma \right) \tau} = 1,
    \label{lambert1}
\end{equation}
after which the substitution
\begin{equation}
u \equiv -\left(\omega_0- \gamma \right) \tau + \frac{\gamma-\omega_0}{\omega_0}
\label{subst}
\end{equation}
allows us to express \eqref{lambert1} in the form called the Lambert equation \cite{Corless}
\begin{equation}
    u e^u = y
    \label{lambert2}
\end{equation}
with the shorthand notation
\begin{equation}
y \equiv \frac{\gamma-\omega_0}{ \omega e^{\frac{\omega_0-\gamma}{\omega_0}}}.
\end{equation}
We note here that $-1/e<y<0$, since $0<\gamma<\omega_0$ in the underdamped regime. In this case, the Lambert equation \eqref{lambert2} has two solutions for the unknown $u$ \cite{Corless}:
\begin{align}
    u_1 &= W_0(y), \ u_2 = W_{-1}(y),
\end{align}
where $W_0$ and $W_{-1}$ are two branches of the Lambert W function. We can write this compactly as $u_{1,2}=W_{0,-1}(y)$.
By inserting the solutions $u_1$ and $u_2$ back into the substitution \eqref{subst}, we arrive at two solutions for crossing times $\tau_1$ and $\tau_2$, as written in \eqref{tau12}:
\begin{equation}
\tau_{1,2}=-\frac{1}{\omega_0-\gamma}W_{0,-1}\left(-\frac{\omega_0-\gamma}{\omega e^{\frac{\omega_0-\gamma}{\omega_0}}}\right)-\frac{1}{\omega_0}.
\end{equation}

\section{Energy ratio \eqref{E} as a function of $\gamma$ at different time instants $t$}
\label{globalmin}

\renewcommand\thefigure{\thesection\arabic{figure}}
\setcounter{figure}{0}    

\begin{figure}[H]
\begin{center}
\begin{tikzpicture}
        \begin{axis}[
        width=0.485\textwidth,
        height=0.4\textwidth,
        xmin = 0,
        xmax = 1.5,
        ymin = -14.5,
        ymax = 1,
        xtick={0,0.2,...,1.4},
        ytick={0,-2,...,-14},
        every tick label/.append style={font=\small},
        ylabel near ticks,
        xlabel near ticks,
        xlabel = {\small $\gamma\,[\omega_0]$},
        ylabel = {\small $\log_{10}\left(E(\gamma,t)/E_0\right)$},
        legend entries = {}
        ]
        \addplot [densely dotted, domain=0:0.999,samples=200,thick, black](\x,{log10(exp(-2*\x*4)*((cos(deg(sqrt(1-\x^2)*4)))^2+\x*sin(deg(2*sqrt(1-\x^2)*4))/sqrt(1-\x^2)+(1+\x^2)*(sin(deg(sqrt(1-\x^2)*4)))^2/(1-\x^2)))});
        \addplot [densely dotted, domain=0:0.999,samples=200,thick, black](\x,{log10(exp(-2*\x*6)*((cos(deg(sqrt(1-\x^2)*6)))^2+\x*sin(deg(2*sqrt(1-\x^2)*6))/sqrt(1-\x^2)+(1+\x^2)*(sin(deg(sqrt(1-\x^2)*6)))^2/(1-\x^2)))});
        \addplot [densely dotted, domain=0:0.999,samples=200,thick, black](\x,{log10(exp(-2*\x*9)*((cos(deg(sqrt(1-\x^2)*9)))^2+\x*sin(deg(2*sqrt(1-\x^2)*9))/sqrt(1-\x^2)+(1+\x^2)*(sin(deg(sqrt(1-\x^2)*9)))^2/(1-\x^2)))});
        \addplot [densely dotted, domain=0:0.999,samples=200,thick, black](\x,{log10(exp(-2*\x*12)*((cos(deg(sqrt(1-\x^2)*12)))^2+\x*sin(deg(2*sqrt(1-\x^2)*12))/sqrt(1-\x^2)+(1+\x^2)*(sin(deg(sqrt(1-\x^2)*12)))^2/(1-\x^2)))});
        \addplot [densely dotted, domain=0:0.999,samples=200,thick, black](\x,{log10(exp(-2*\x*15)*((cos(deg(sqrt(1-\x^2)*15)))^2+\x*sin(deg(2*sqrt(1-\x^2)*15))/sqrt(1-\x^2)+(1+\x^2)*(sin(deg(sqrt(1-\x^2)*15)))^2/(1-\x^2)))});
        \addplot [densely dotted, domain=1.001:1.5,samples=100,thick, black](\x,{log10(exp(-2*\x*4)*((cosh((sqrt(\x^2-1)*4)))^2+\x*sinh((2*sqrt(\x^2-1)*4))/sqrt(\x^2-1)+(1+\x^2)*(sinh((sqrt(\x^2-1)*4)))^2/(\x^2-1)))});
        \addplot [densely dotted, domain=1.001:1.5,samples=100,thick, black](\x,{log10(exp(-2*\x*6)*((cosh((sqrt(\x^2-1)*6)))^2+\x*sinh((2*sqrt(\x^2-1)*6))/sqrt(\x^2-1)+(1+\x^2)*(sinh((sqrt(\x^2-1)*6)))^2/(\x^2-1)))});
        \addplot [densely dotted, domain=1.001:1.5,samples=100,thick, black](\x,{log10(exp(-2*\x*9)*((cosh((sqrt(\x^2-1)*9)))^2+\x*sinh((2*sqrt(\x^2-1)*9))/sqrt(\x^2-1)+(1+\x^2)*(sinh((sqrt(\x^2-1)*9)))^2/(\x^2-1)))});
        \addplot [densely dotted, domain=1.001:1.5,samples=100,thick, black](\x,{log10(exp(-2*\x*12)*((cosh((sqrt(\x^2-1)*12)))^2+\x*sinh((2*sqrt(\x^2-1)*12))/sqrt(\x^2-1)+(1+\x^2)*(sinh((sqrt(\x^2-1)*12)))^2/(\x^2-1)))});
        \addplot [densely dotted, domain=1.001:1.5,samples=100,thick, black](\x,{log10(exp(-2*\x*15)*((cosh((sqrt(\x^2-1)*15)))^2+\x*sinh((2*sqrt(\x^2-1)*15))/sqrt(\x^2-1)+(1+\x^2)*(sinh((sqrt(\x^2-1)*15)))^2/(\x^2-1)))});
        \node[fill, black, circle,inner sep=0.85pt,label={}] at (0.786,-2.729){}; 
        \node[fill, black, circle,inner sep=0.85pt,label={}] at (0.895,-4.659){}; 
        \node[fill, black, circle,inner sep=0.85pt,label={}] at (0.9495,-7.4264){};
        \node[fill, black, circle,inner sep=0.85pt,label={}] at (0.971,-10.012){};
        \node[fill, black, circle,inner sep=0.85pt,label={}] at (0.981,-12.625){};
        \node[] at (axis cs: 0.55,-0.65) {\footnotesize $t=4\omega_0^{-1}$};
        \node[] at (axis cs: 0.55,-10) {\footnotesize $t=15\omega_0^{-1}$};
        \draw[->,>=stealth] (axis cs: 0.5,-1) -- (axis cs: 0.5,-9.5);
        
        \end{axis}  
    \end{tikzpicture}
\end{center}
\caption{The base 10 logarithm of the energy ratio \eqref{E} as a function of $\gamma$ at instants $t=\lbrace4\omega_0^{-1},6\omega_0^{-1}, 9\omega_0^{-1}, 12\omega_0^{-1}, 15\omega_0^{-1}\rbrace$ (dotted black curves, ordered from top to bottom, as indicated by the arrow). A small black dot on each of the curves indicates the value for $\gamma$ with which the oscillator at that moment comes to the equilibrium position for the first time.}
\label{fig:FigB1}
\end{figure}
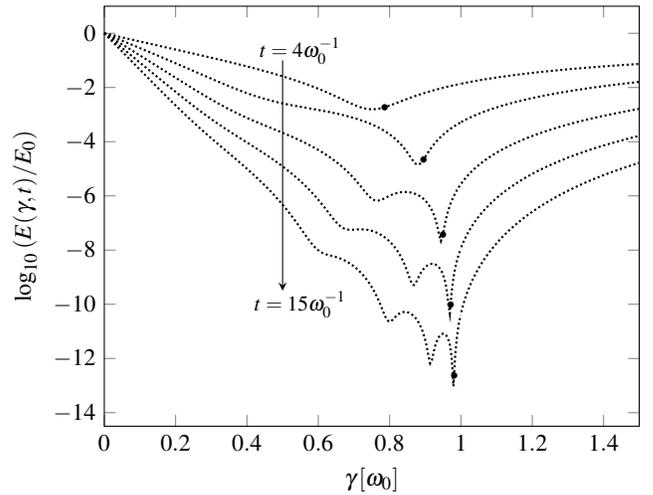

In Fig.\,\ref{fig:FigB1} we show the base 10 logarithm of the energy ratio \eqref{E} as a function of $\gamma\in[0,1.5\omega_0]$ for five time instants. We see that as time increases, new local minima are formed, but the global minimum persists and its position approaches the critical damping value $\gamma=\omega_0$. The black dot on each of the curves indicates the value for $\gamma$ for which the system is at the equilibrium position for the first time at that instant. We see that the subset of the damping coefficients, for which the energy is lower than the energy for this $\gamma$, decreases in range as time increases. Thus, the difference between $\gamma_1$ and $\gamma_{opt}$ (from subsection \ref{subsection:optimum}) decreases as we optimize towards lower and lower energies. Behavior for $t>15\omega_0^{-1}$ is qualitatively the same.    

 \end{appendix}

\nocite{*} 

\providecommand{\noopsort}[1]{}\providecommand{\singleletter}[1]{#1}%
%

\clearpage
\begin{center}
\textbf{\large Supplementary material for "Damped harmonic oscillator revisited: the fastest route to equilibrium"}
\end{center}
\setcounter{equation}{0}
\setcounter{figure}{0}
\setcounter{table}{0}
\setcounter{page}{1}
\setcounter{section}{0}
\counterwithout{equation}{section}
\makeatletter
\renewcommand{\theequation}{S\arabic{equation}}
\renewcommand{\thefigure}{S\arabic{figure}}
\renewcommand{\thetable}{S\Roman{table}}
\renewcommand{\thesection}{S\Roman{section}}
\renewcommand{\thepage}{S\arabic{page}}

%







This supplementary material contains additional equations, figures and tables. Such items are numbered with an uppercase "S" to identify them as supplementary. For example, the first figure in the supplemental document is "Fig. S1"; the second table, "Table SII"; etc. Equations, figures and tables mentioned in this supplementary text, which are not marked with "S", refer to those same items in the main text.

\section{Proof of $x_{od}(t) \geq x_{c}(t)$ when starting from the same initial displacement at rest}
\label{AppencixODCR}
Let us look at the solutions for the critically damped and the overdamped oscillator, both starting from rest with the same displacement from equilibrium:
\begin{equation}
x_c(t)=x_0\left(1+\omega_0 t\right)e^{-\omega_0 t}\,,
\label{cs}
\end{equation} 
\begin{equation}
x_{od}(t)=x_0 e^{-\gamma t}\left(\frac{\gamma+\alpha}{2\alpha}e^{\alpha t}-\frac{\gamma-\alpha}{2\alpha}e^{-\alpha t}\right)\,,
\end{equation}  
where $\alpha = \sqrt{\gamma^2-\omega_0^2}$ and $\gamma > \omega_0$. Here we use the fact that both $x_c(t)$ and $x_{od}(t)$ are entire functions so we can use their Taylor expansions around any point to compare them. First, we recast the overdamped solution in a sum of hyperbolic functions, making all of the coefficients of the Taylor expansions positive for easier comparison:
\begin{equation}
x_{od}(t)=x_0 e^{-\gamma t}\left(\frac{\gamma}{\alpha}\sinh(\alpha t) + \cosh(\alpha t)\right)\,.
\label{hint}
\end{equation}
The Taylor expansion of the overdamped case around $\alpha t = 0$ is equal to:
\begin{eqnarray}
x_{od}(t)&=&x_0 e^{-\gamma t}\sum\limits_{n=0}^{\infty}\left(\frac{\gamma}{\alpha}\frac{(\alpha t)^{2n+1}}{(2n+1)!} + \frac{(\alpha t)^{2n}}{(2n)!}\right)\nonumber \\
&=&x_0 e^{-\gamma t}\left(\frac{\gamma}{\alpha}\left(\frac{\alpha t}{1!} + \frac{(\alpha t)^3}{3!} + ...\right) + 1 + \frac{(\alpha t)^2}{2!} + ... \right) \nonumber \\
&=&x_0 e^{-\gamma t}\left(1+ \frac{\gamma t^1}{1!}+\frac{\alpha^2 t^2}{2!} + \frac{\gamma \alpha^2 t^3}{3!} +  ... \right)\,. \qquad
\end{eqnarray}
Our goal is to compare the displacements of the critically damped and the overdamped case. We claim that for the selected initial conditions the displacement of the overdamped case will always be greater or equal than the displacement of the critically damped case, $x_{od}(t) \geq x_{c}(t)$. Dividing both sides of this inequality with $x_0$ and multiplying by $e^{\gamma t}$ leaves the inequality sign unchanged since both of these factors are positive. We thus have, using the already calculated Taylor expansion of the left-hand side:
\begin{equation}
\left(1+ \frac{\gamma t^1}{1!}+\frac{\alpha^2 t^2}{2!} + \frac{\gamma \alpha^2 t^3}{3!} + ... \right) \geq (1+\omega_0 t)e^{(\gamma - \omega_0)t} \,.
\label{A5}
\end{equation}
The Taylor expansion of the right-hand side of (\ref{A5}) around $(\gamma-\omega_0) t=0$ is equal to:
\begin{eqnarray} 
1+\sum\limits_{n=1}^{\infty}\frac{(\gamma-\omega_0)^{n-1}(\gamma+(n-1)\omega_0)t^n}{n!}\,.\qquad
\end{eqnarray}
We can now compare the left- and right-hand sides of (\ref{A5}) by comparing the corresponding coefficients of $t^n$. First, note that the coefficients of $t^0$ and $t^1$ coincide on both sides. Next, we can see that the left-hand side expansion has a different coefficient depending on the parity of n. We start by looking at the case when $n$ is even and set it to $n=2k$, with $k \in \mathbb{N}$. Comparing the powers of $t^{2k}$ gives:
\begin{eqnarray}
\frac{\alpha^{2k}}{(2k)!} \geq \frac{(\gamma-\omega_0)^{2k-1}(\gamma+(2k-1)\omega_0)}{(2k)!}.
\end{eqnarray}
Cancelling $(2k)!$ and expanding $\alpha = \sqrt{\gamma^2 - \omega_0^2}$ we get:
\begin{eqnarray}
(\gamma^2-\omega_0^2)^k &\geq& (\gamma + (2k-1)\omega_0)(\gamma-\omega_0)^{2k-1} \nonumber \\
\left[(\gamma - \omega_0)(\gamma+\omega_0)\right]^k &\geq& (\gamma + (2k-1)\omega_0)(\gamma-\omega_0)^{2k-1}\,. \qquad
\end{eqnarray}
Cancelling $(\gamma-\omega_0)^k$ won't change the inequality sign since $\gamma > \omega_0$. We are therefore left with:
\begin{eqnarray}
(\gamma + \omega_0)^k \geq (\gamma + (2k-1)\omega_0)(\gamma-\omega_0)^{k-1}\,.
\end{eqnarray}
Simplyfing the right-hand side further we get:
\begin{eqnarray}
(\gamma + \omega_0)^k &\geq& (\gamma + 2k\omega_0 - \omega_0)(\gamma - \omega_0)^{k-1} \nonumber \\
&\geq& (\gamma - \omega_0)^k + 2k\omega_0(\gamma - \omega_0)^{k-1} \,.
\label{A10}
\end{eqnarray}
The left-hand side can be written as $((\gamma - \omega_0)+ 2\omega_0)^k$ and expanded using the binomial theorem:
\begin{eqnarray}
((\gamma - \omega_0)+ 2\omega_0)^k &=& \sum\limits_{i=0}^k\binom{k}{i}  (\gamma - \omega_0)^{k-i}(2\omega_0)^i \nonumber \\
&=& (\gamma - \omega_0)^k + k(\gamma - \omega_0)^{k-1}(2\omega_0) \nonumber \\
&+& \sum\limits_{i=2}^k\binom{k}{i}  (\gamma - \omega_0)^{k-i}(2\omega_0)^i \,.
\end{eqnarray}
We can see that the first two terms of the binomial expansion, which have been written explicitly, equal the right-hand side of (\ref{A10}). All the other terms of the expansion are positive, making the left-hand side of (\ref{A10}) strictly larger that the right-hand side of (\ref{A10}) whenever $k > 1$ and the two sides are equal in case $k=1$.

Next, we look at the case when n is odd and set it to $n=2k+1$, with $k \in \mathbb{N}$. Comparing the powers of $t^{2k+1}$ gives:
\begin{eqnarray}
\frac{\gamma\alpha^{2k}}{(2k+1)!} > \frac{(\gamma-\omega_0)^{2k}(\gamma+2k\omega_0)}{(2k+1)!}.
\end{eqnarray}
Note that in this case we assume that he left-hand side of the equation is strictly larger than the right-hand side, which we now demonstrate. As earlier, we cancel $(2k+1)!$ and expand $\alpha = \sqrt{\gamma^2 - \omega_0^2}$ to get:
\begin{eqnarray}
\gamma(\gamma^2-\omega_0^2)^k &>& (\gamma + 2k\omega_0)(\gamma-\omega_0)^{2k} \nonumber \\
\gamma\left[(\gamma - \omega_0)(\gamma+\omega_0)\right]^k &>& (\gamma + 2k\omega_0)(\gamma-\omega_0)^{2k}\,.
\end{eqnarray}
Next, we cancel $(\gamma-\omega_0)^k$ and are left with
\begin{eqnarray}
\gamma(\gamma + \omega_0)^k > (\gamma + 2k\omega_0)(\gamma-\omega_0)^{k}\,.
\label{A14}
\end{eqnarray}
The left-hand side can be written as $\gamma((\gamma - \omega_0)+ 2\omega_0)^k$ and expanded using the binomial theorem:
\begin{eqnarray}
\gamma((\gamma - \omega_0)+ 2\omega_0)^k &=& \gamma\sum\limits_{i=0}^k\binom{k}{i}  (\gamma - \omega_0)^{k-i}(2\omega_0)^i \nonumber \\
&=& \gamma(\gamma - \omega_0)^k + \gamma k(\gamma - \omega_0)^{k-1}(2\omega_0) \nonumber \\
&+& \gamma \sum\limits_{i=2}^k\binom{k}{i}  (\gamma - \omega_0)^{k-i}(2\omega_0)^i \,.
\end{eqnarray}
The second term of the expansion can be rewritten as
\begin{eqnarray}
\gamma k(\gamma - \omega_0)^{k-1}(2\omega_0)&=& (\gamma-\omega_0+\omega_0) k(\gamma - \omega_0)^{k-1}(2\omega_0) \nonumber\\
&=&2k\omega_0(\gamma-\omega_0)^k + 2k\omega_0^2(\gamma-\omega_0)^{k-1}\,. \nonumber
\end{eqnarray}
This makes the entire left hand side equal to:
\begin{eqnarray}
\gamma(\gamma + \omega_0)^k &=& \gamma(\gamma - \omega_0)^k + 2k\omega_0(\gamma-\omega_0)^k \nonumber \\
&+& 2k\omega_0^2(\gamma-\omega_0)^{k-1} + \gamma \sum\limits_{i=2}^k\binom{k}{i}  (\gamma - \omega_0)^{k-i}(2\omega_0)^i\,. \nonumber
\end{eqnarray}
Since the first two terms coincide with the right-hand side of (\ref{A14}) and all the other terms are positive, we conclude that the inequality holds true. This is also true in the special case when $k=1$ since then the left-hand side equals
\begin{eqnarray}
\gamma(\gamma + \omega_0)^1 = \gamma(\gamma - \omega_0)^1 + 2\omega_0(\gamma-\omega_0)^1 + 2\omega_0^2\,,
\end{eqnarray}
which is clearly larger that the right-hand side in this special case. In conclusion, the left-hand side of (\ref{A5}) is in all cases equal or larger than the right-hand side of (\ref{A5}) and therefore the statement that $x_{od}(t) \geq x_{c}(t)$ is always true with these initial conditions, with the equality holding true only for $t=0$.


\section{Initial condition: zero elongation and maximum velocity}
\label{subsection:v0}

\begin{figure}[h!t!]
\begin{center}
\begin{tikzpicture}
        \begin{axis}[
        width=0.485\textwidth,
        height=0.3\textwidth,
        xmin = 0,
        xmax = 7,
        ymin = 0,
        ymax = 0.45,
        xtick={0,1,...,6,7},
        ytick={0,0.1,0.2,0.3,0.4},
        every tick label/.append style={font=\small},
        ylabel near ticks,
        xlabel near ticks,
        xlabel = {\small $t\,[\omega_0^{-1}]$},
        ylabel = {\small $\omega_0 x(t)/v_0$},
        legend entries = {\scriptsize $\gamma=0.9\omega_0$\\\scriptsize $\gamma=\omega_0$\\\scriptsize $\gamma=1.1\omega_0$\\\scriptsize $\gamma=5\omega_0$\\}
        ]
        \addplot [domain=0:7,samples=200,thick, red](\x,{2.294*exp(-0.9*\x)*sin(deg(0.436*\x)});
        \addplot [domain=0:7,samples=200,thick, dashed](\x,{\x*exp(-\x)});
        \addplot [domain=0:7,samples=200,thick, blue, densely dotted](\x,{2.182*exp(-1.1*\x)*(1-exp(-0.917*\x))/(2*exp(-0.458*\x))});
        \addplot [domain=0:7,samples=200,thick, red, densely dotted](\x,{0.204*exp(-5*\x)*(1-exp(-9.798*\x))/(2*exp(-4.899*\x))});
        
        \node[] at (axis cs: 4.2,0.35) {(a)};
        
        \end{axis}  
    \end{tikzpicture}
\end{center}
\begin{center}
\hskip -4mm
\begin{tikzpicture}
        \begin{axis}[
        width=0.485\textwidth,
        height=0.3\textwidth,
        xmin = 0,
        xmax = 7,
        ymin = 0,
        ymax = 0.01,
        xtick={0,1,...,6,7},
        yticklabel style={
        /pgf/number format/fixed,
        /pgf/number format/precision=3
        },
        scaled y ticks=false,
        ytick={0,0.002,0.004,0.006,0.008,0.01},
        every tick label/.append style={font=\small},
        ylabel near ticks,
        xlabel near ticks,
        xlabel = {\small $t\,[\omega_0^{-1}]$},
        ylabel = {\small $\omega_0 x(t)/v_0$},
        legend entries = {\scriptsize $\gamma=0.9\omega_0$\\\scriptsize $\gamma=\omega_0$\\\scriptsize $\gamma=100\omega_0$\\\scriptsize $\gamma=200\omega_0$\\},
        legend style={at={(0.02,0.98)},anchor=north west}
        ]
        \addplot [domain=0:7,samples=200,thick, red](\x,{2.294*exp(-0.9*\x)*sin(deg(0.436*\x)});
        \addplot [domain=0:7,samples=200,thick, dashed](\x,{\x*exp(-\x)});
        \addplot [domain=0:7,samples=200,thick, blue, densely dotted](\x,{0.01*exp(-100*\x)*(1-exp(-200*\x))/(2*exp(-100*\x))});
        \addplot [domain=0:7,samples=200,thick, red, densely dotted](\x,{0.005*exp(-200*\x)*(1-exp(-400*\x))/(2*exp(-200*\x))});
        
        \node[] at (axis cs: 4.2,0.0075) {(b)};
        
        \end{axis}  
    \end{tikzpicture}
\end{center}
\caption{(a) The underdamped solution with $\gamma=0.9\omega_0$ (red solid curve), the critically damped solution (black dashed curve), and the overdamped solutions with $\gamma=1.1\omega_0$ (blue dotted curve) and $\gamma=5\omega_0$ (red dotted curve). All the solutions are scaled by the same factor making them dimensionless. (b) The underdamped solution (red solid curve), the critically damped solution (black dashed curve), and overdamped solutions with $\gamma=100\omega_0$ (blue dotted curve) and $\gamma=200\omega_0$ (red dotted curve). The overdamped solutions have a step-like behaviour when $\gamma\gg\omega_0$.}
\label{fig:Fig7S}
\end{figure}
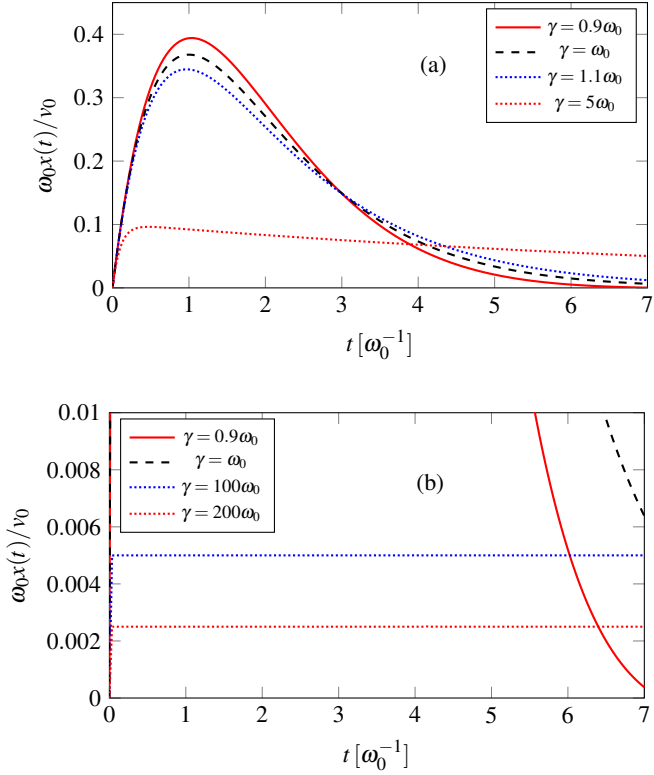

We next focus on the case with initial conditions $x_0=0$ and $v_0>0$. Here we present the solutions for all three oscillator regimes with the given initial condition taken into account
\begin{eqnarray}
\label{uds2}
x_{ud}(t)&=&\frac{v_0}{\omega}e^{-\gamma t}\sin\left(\omega t\right) \,, \\
\label{cs2}
x_c(t)&=&v_0 t e^{-\omega_0 t}\,,\\
x_{od}(t)&=&\frac{v_0}{\alpha}e^{-\gamma t}\sinh(\alpha t) =\frac{v_0}{\alpha}e^{-\gamma t}\frac{1-e^{-2\alpha t}}{2e^{-\alpha t}} \,,
\label{ods2}
\end{eqnarray}
where the already introduced notation was used. In Fig.\,\ref{fig:Fig7S}(a) we show the underdamped solution $x_{ud}(t)$ with $\gamma=0.9\omega_0$ as a red solid curve, the critically damped solution $x_c(t)$ as a black dashed curve, and the overdamped solutions $x_{od}(t)$ with $\gamma=1.1\omega_0$ as the blue dotted curve and $\gamma=5\omega_0$ as the red dotted curve. All the solutions are scaled by the same factor making them dimensionless. We see that at the end of the given time range the underdamped solution is closest to the equilibrium. Looking at the overdamped curves in Fig.\,\ref{fig:Fig7S}(a) one might think that it is possible to choose a very large damping coefficient $\gamma$ and reach the equilibrium sooner than any of the underdamped and/or critically damped oscillators. We argue here that this is not the case. In Fig.\,\ref{fig:Fig7S}(b) we show the underdamped and the critically damped solution, just as in Fig.\,\ref{fig:Fig7S}(a), as well as overdamped solutions with $\gamma=100\omega_0$ as a blue dotted curve and $\gamma=200\omega_0$ as a red dotted curve. We see that for $\gamma\gg\omega_0$, the overdamped solution has a step-like behaviour which can be explained by an analytical approximation as follows. For $\gamma\gg\omega_0$ we have 
\begin{equation}
\alpha=\sqrt{\gamma^2-\omega_0^2}=\gamma\sqrt{1-\left(\frac{\omega_0}{\gamma}\right)^2}\approx\gamma \,.
\label{biggamma}
\end{equation}
Using \eqref{biggamma} and \eqref{ods2} we get
\begin{equation}
x_{od}(t)\approx\frac{v_0}{2\gamma}\left(1-e^{-2\gamma t}\right).
\label{ods2approx}
\end{equation}   
From \eqref{ods2approx} we see that in this limit the overdamped solution rapidly rises form zero at $t=0$ to $v_0/(2\gamma)$ for $t \gtrsim 1/\gamma$, which is in perfect agreement with the results shown in Fig.\,\ref{fig:Fig7S}(b). In conclusion, selecting a very large $\gamma$ (comapared to $\omega_0$) makes the oscillator reach a practically infinitesimal maximal elongation very quickly before it very slowly starts to return to the equilibrium position. Therefore, similarly as in the main text, we exclude the overdamped regime from further analysis in this subsection.  

\begin{figure}[h!t!b!]
\begin{center}
\begin{tikzpicture}
        \begin{axis}[
        width=0.485\textwidth,
        height=0.3\textwidth,
        xmin = 0,
        xmax = 10,
        ymin = -9,
        ymax = 1,
        xtick={0,1,...,9,10},
        ytick={-8,-6,...,-2,0},
        every tick label/.append style={font=\small},
        ylabel near ticks,
        xlabel near ticks,
        xlabel = {\small $t\,[\omega_0^{-1}]$},
        ylabel = {\small $\log_{10} \left(E(t)/E_0\right)$},
        legend entries = {\scriptsize $\gamma=0.85\omega_0$\\\scriptsize $\gamma=0.9\omega_0$\\\scriptsize $\gamma=0.95\omega_0$\\\scriptsize $\gamma=\omega_0$\\}
        ]
        \addplot [domain=0:10,samples=200,thick, dash dot, green](\x,{-0.738*\x+log10(3.6-2.6*cos(deg(1.054*\x))-1.613*sin(deg(1.054*\x)))});
        \addplot [domain=0:10,samples=200,thick, red](\x,{-0.782*\x+log10(5.262-4.263*cos(deg(0.872*\x))-2.065*sin(deg(0.872*\x)))});
        \addplot [domain=0:10,samples=200,thick, blue, densely dotted](\x,{-0.825*\x+log10(10.256-9.256*cos(deg(0.624*\x))-3.042*sin(deg(0.624*\x)))});
        \addplot [domain=0:10,samples=100,thick,dashed, black](\x,{log10((1-\x)^2+\x^2)-0.869*\x});

        \addplot [domain=0:10,samples=100,dashed, black](\x,{-6});
        \node[] at (axis cs: 2,-5.4) {\small $E(t) = 10^{-6}E_0$};
        
        \end{axis}  
    \end{tikzpicture}
\end{center}
\caption{Base 10 logarithm of the ratio (9), for underdamped oscillators with $\gamma=0.85\omega_0$ (green dash-dotted curve), $\gamma=0.9\omega_0$ (red solid curve), $\gamma=0.95\omega_0$ (blue dotted curve) and the critically damped oscillator (black dashed curve). A line of reference energy $E(t) = 10^{-6}E_0$ has been added for easier comparison of curves.}
\label{fig:Fig8}
\end{figure}
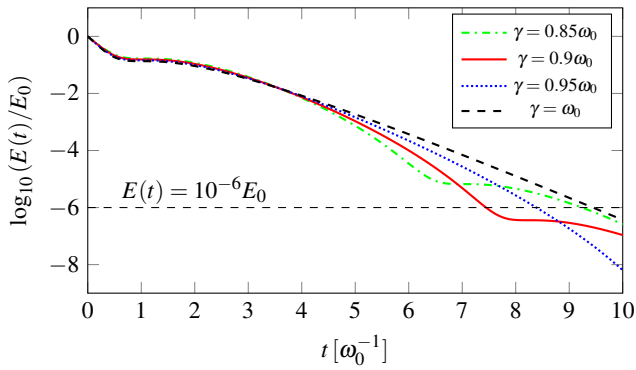

In Fig.\,\ref{fig:Fig8} we show the base 10 logarithm of the ratio (9), for underdamped oscillators with $\gamma=0.85\omega_0$ as the green dash-dotted curve, $\gamma=0.9\omega_0$ as the red solid curve, $\gamma=0.95\omega_0$ as the blue dotted curve and the critically damped oscillator as the black dashed curve. We note that with these initial conditions the behaviour of the oscillators is qualitatively very similar to the one shown in Fig. 3(a). Thus, we omit the finer details and proceed with the optimisation as described in the main text. 

The underdamped oscillator, with given initial conditions, reaches the equilibrium position for the $n$-th time at moments given by
\begin{equation}
\tau_{A_n}=\frac{n\pi}{\omega},
\label{tauA2}
\end{equation}
where $n\in\mathbb{N}$. We define the moment the system has settled down in the equilibrium as the moment when its energy is equal to $10^{-\delta}E_0$, with some $\delta>0$. The underdamped coefficient $\gamma_1$ is then found as the one for which the system reaches this energy level at the moment given by $\tau_{A_1}$. Since at this moment the total energy of the oscillator is equal to its kinetic energy, we can write the equilibrium condition as
\begin{equation}
\frac{E_{ud}(\tau_{A_1})}{E_0}=\frac{\dot x_{ud}(\tau_{A_1})^2}{v_0^2}=10^{-\delta}\,.
\label{condition11}
\end{equation}
By using \eqref{tauA2} with $n=1$ and the time derivative of \eqref{uds2}, we get
\begin{equation}
\exp\left(-\frac{2\gamma\pi}{\sqrt{\omega_0^2-\gamma^2}}\right)=10^{-\delta}.
\label{condition22}
\end{equation}
In Table SI. we show results obtained for $\gamma_1$ from this condition. Considering the energy ratio (9) as a function of $\gamma$ at $\tau_{A_1}$, it is possible to obtain $\gamma_{opt}$ in qualitatively the same way as in the main text.

\begingroup
\setlength{\tabcolsep}{5.2pt} 
\renewcommand{\arraystretch}{1.5} 
\begin{table}[h!t!]
\begin{tabular}{ |c|c|c|c|c| } 
 \hline
 $E(t)/E_0$ & $\gamma_1 \left[ \omega_0 \right]$ & $\tau_{A_1} [ \omega_0^{-1}]$ & $\tau_{c}[ \omega_0^{-1}]$ & $\left(\tau_{c} - \tau_{A_1} \right)/\tau_{c} \left[ \% \right]$ \\ 
 \thickhline
 $10^{-4}$ & \ $0.8261$ \ & $5.57$ & $6.79$ & $17.97$ \\ 
 \hline
 $10^{-6}$ & \ $0.9103$ \ & $7.59$ & $9.45$ & $19.68$ \\ 
 \hline
 $10^{-8}$ & \ $0.9465$ \ & $9.74$ & $12.00$ & $18.83$ \\ 
 \hline
 $10^{-10}$ & \ $0.9647$ \ & $11.93$ & $14.50$ & $17.72$ \\ 
 \hline
 $10^{-12}$ & \ $0.9751$ \ & $14.17$ & $16.96$ & $16.45$ \\ 
 \hline
 $10^{-14}$ & \ $0.9815$ \ & $16.41$ & $19.40$ & $15.41$ \\ 
 \hline
 $10^{-16}$ & \ $0.9858$ \ & $18.71$ & $21.83$ & $14.29$ \\ 
 \hline
 $10^{-18}$ & \ $0.9887$ \ & $20.96$ & $24.24$ & $13.53$ \\ 
 \hline
\end{tabular}
\caption{The underdamped coefficients $\gamma_1$ (col.2) obtained from condition \eqref{condition22} for various choices of $\delta$ (col.1). The last three columns show the moments in time when the condition \eqref{condition22} is satisfied for the $\gamma_1$ (col.3), the time when the energy of a critically damped oscillator drops to the same level (col.4) and the relative time advantage of the underdamped oscillator compared to the critically damped oscillator (col.5).}
\end{table}
\endgroup

\section{Same sign initial elongation and velocity}
\label{SameXV}

The case in which the initial conditions are such that $x_0 \neq 0$ and $v_0 \neq 0$, but are of the same sign is a direct combination of cases already explored in main text and in previous subsection. Therefore, the conclusions drawn there also hold true here and we can always find an optimal underdamped coefficient such that the energy of the oscillator with this $\gamma$ drops to a level under a preset threshold when arriving in the equilibrium position for the first (or second, third, etc.) time. The relative time advantage of the underdamped oscillator as compared to the critically damped oscillator will depend on the initial conditions as well as the preset value of energy in an algebraically non-trivial way, but will nevertheless be greater than zero for the best-case optimization choice. Therefore, we omit the finer details in this section and conclude that our findings are still valid.

\section{Opposite sign initial elongation and velocity} 
\label{OppXV}

Finally, we focus on the case with initial conditions $x_0>0$ and $v_0<0$ (or vice-versa, which can be accounted for by a coordinate sign flip). As we will see in what follows, for these initial conditions, the behaviour of the oscillator in the underdamped regime is qualitatively the same as for already analysed initial conditions, while the critically damped and the overdamped regimes show some new qualities. Therefore, here we repeat the solutions for the critically damped and the overdamped oscillators, (3) and (4), with $v_0=-|v_0|$ taken into account
\begin{equation}
x_c(t)=\left(x_0+\left(\omega_0 x_0 -|v_0|\right)t\right)e^{-\omega_0 t},
\label{cds3}
\end{equation} 
\begin{align}
x_{od}(t)=e^{-\gamma t}&\biggl(\frac{(\gamma+\alpha)x_0-|v_0|}{2\alpha}e^{\alpha t}\nonumber \\
&-\frac{(\gamma-\alpha)x_0-|v_0|}{2\alpha}e^{-\alpha t}\biggr)\,.
\label{ods3}
\end{align}
Depending on the relationship between the initial kinetic potential energies, we analyse three cases. 

\subsection{The initial kinetic energy is smaller than the initial potential energy}
\label{subsubsection:Ekless} 

The condition that the initial kinetic energy is smaller than the initial potential energy can be recast in the form $|v_0|<\omega_0 x_0$. If this condition is true, the critically damped solution \eqref{cds3} has no zeroes, so it asymptotically approaches the equilibrium position without ever overshooting it. The overdamped solution \eqref{ods3} approaches the equilibrium position more slowly than the critically damped solution due to a higher damping coefficient. Thus, we can conclude that the overdamped solution is further away from the equilibrium, when compared to the critically damped solution, for any $t>0$. Therefore, for the selected initial condition the overdamped oscillator also never overshoots the equilibrium position.

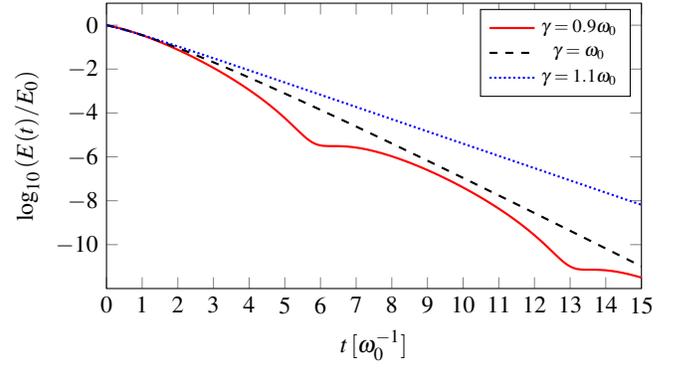
\begin{figure}[h!t!]
\begin{center}
\begin{tikzpicture}
        \begin{axis}[
        width=0.485\textwidth,
        height=0.3\textwidth,
        xmin = 0,
        xmax = 15,
        ymin = -12,
        ymax = 1,
        xtick={0,1,...,14,15},
        ytick={-10,-8,...,-2,0},
        every tick label/.append style={font=\small},
        ylabel near ticks,
        xlabel near ticks,
        xlabel = {\small $t\,[\omega_0^{-1}]$},
        ylabel = {\small $\log_{10} \left(E(t)/E_0\right)$},
        legend entries = {\scriptsize $\gamma=0.9\omega_0$\\\scriptsize $\gamma=\omega_0$\\\scriptsize $\gamma=1.1\omega_0$\\}
        ]
        \addplot [domain=0:15,samples=200,thick, red](\x,{log10(0.8*((1.25-0.9)*exp(-1.8*\x)*(5.262+4.263*cos(deg(0.872*\x-1.485))+2.065*sin(deg(0.872*\x-1.485)))))});
        \addplot [domain=0:15,samples=200,thick, dashed](\x,{log10(0.2*exp(-2*\x)*((1+\x)^2+(2+\x)^2))});
        \addplot [domain=0:15,samples=200,thick, densely dotted, blue](\x,{log10(0.2*exp(-2.2*\x)*(7.2*sinh(0.916*\x)-2.4*(4*2.183-5*2.4)*cosh(0.916*\x)-2.183*(5*2.183-4*2.4)))});
        \end{axis}  
    \end{tikzpicture}
\end{center}
\caption{The base 10 logarithm of the ratio (9) for the underdamped oscillator with $\gamma=0.9\omega_0$ (solid red curve), the critically damped oscillator (black dashed curve) and the overdamped oscillator with $\gamma=1.1\omega_0$ (blue dotted curve). The curves are drawn with the condition $|v_0|=\omega_0x_0/2$.}
\label{fig:Fig9}
\end{figure}

In Fig.\,\ref{fig:Fig9} we show the base 10 logarithm of the ratio (9) for the underdamped oscillator with $\gamma=0.9\omega_0$ as the solid red curve, the critically damped oscillator as the black dashed curve and the overdamped oscillator with $\gamma=1.1\omega_0$ as the blue dotted curve. To draw the curves, we selected $|v_0|=\omega_0x_0/2$. We see that for this type of initial conditions, the behaviour of the underdamped oscillator compared to the critically damped oscillator is qualitatively the same as already examined and shown in Fig. 3(a) and Fig.\,\ref{fig:Fig8}. Thus, one can perform the optimisation of the underdamped coefficient similarly as before. 

\subsection{The initial kinetic energy is equal to the initial potential energy}
\label{subsubsection:Ekequal}
  
The case when the initial kinetic energy is equal to the initial potential energy can be written as $|v_0|=\omega_0 x_0$. In this case, the linear term in the critically damped solution \eqref{cds3} cancels and we are left with purely exponential behaviour, $x_c(t)=x_0 e^{-\omega_0t}$. Thus, the critically damped solution is again without any zeroes, i.e., it never reaches the equilibrium position. Following the same line of reasoning as in \ref{subsubsection:Ekless} we can conclude that the displacement of the overdamped oscillator \eqref{ods3} is greater than the displacement of the critically damped oscillator for any $t>0$. 

\begin{figure}[h!t!]
\begin{center}
\begin{tikzpicture}
        \begin{axis}[
        width=0.485\textwidth,
        height=0.3\textwidth,
        xmin = 0,
        xmax = 15,
        ymin = -14,
        ymax = 1,
        xtick={0,1,...,14,15},
        ytick={-12,-10,...,-2,0},
        every tick label/.append style={font=\small},
        ylabel near ticks,
        xlabel near ticks,
        xlabel = {\small $t\,[\omega_0^{-1}]$},
        ylabel = {\small $\log_{10} \left(E(t)/E_0\right)$},
        legend entries = {\scriptsize $\gamma=0.9\omega_0$\\\scriptsize $\gamma=\omega_0$\\\scriptsize $\gamma=1.1\omega_0$\\}
        ]
        \addplot [domain=0:15,samples=200,thick, red](\x,{log10(((1-0.9)*exp(-1.8*\x)*(5.262+4.263*cos(deg(0.872*\x+0.225))+2.065*sin(deg(0.872*\x+0.225)))))});
        \addplot [domain=0:15,samples=200,thick, dashed](\x,{log10(exp(-2*\x))});
        \addplot [domain=0:15,samples=200,thick, densely dotted, blue](\x,{log10(exp(-2.2*\x)*(1+1.1*cosh(0.916*\x))/(1+1.1)))});
        \end{axis}  
    \end{tikzpicture}
\end{center}
\caption{The base 10 logarithm of the ratio (9) for an underdamped oscillator with $\gamma=0.9\omega_0$ (solid red curve), the critically damped oscillator (black dashed curve) and the overdamped oscillator with $\gamma=1.1\omega_0$ (blue dotted curve) for the case when the initial kinetic and potential energies are the same.}
\label{fig:Fig10}
\end{figure}
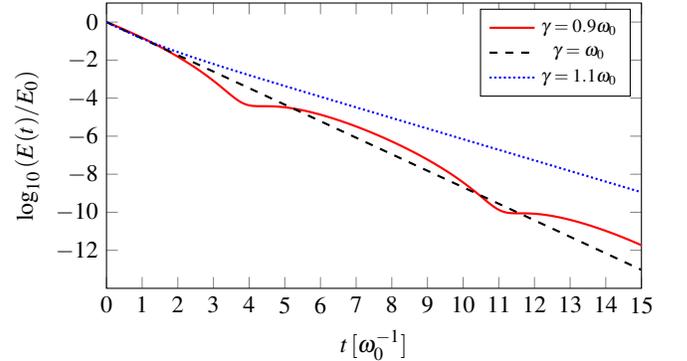

In Fig.\,\ref{fig:Fig10} we show the base 10 logarithm of the ratio (9) for an underdamped oscillator with $\gamma=0.9\omega_0$ as the solid red curve, the critically damped oscillator as the black dashed curve and the overdamped oscillator with $\gamma=1.1\omega_0$ as the blue dotted curve. In this case, we can notice that the underdamped oscillator energy is equal to the energy of the critically damped oscillator for the first time already when $t=5\omega_0^{-1}$. This is much earlier compared to the first crossing over moment $t=24.24\omega_0^{-1}$, shown in Fig. 3(b). Thus, one can in principle perform the optimisation of the underdamped coefficient similarly as before, but with a smaller advantage compared to the critically damped solution.   

\subsection{The initial kinetic energy is greater than the initial potential energy}
\label{Ekgrater}

This final case is realized when $|v_0|>\omega_0x_0$. In contrast to previous cases, the critically damped oscillator solution \eqref{cds3} has a zero at
\begin{equation}
t_{c}=\frac{x_0}{|v_0|-\omega_0x_0} \,.
\label{tc}
\end{equation}
Thus, it overshoots the equilibrium position, reaches a new maximal elongation and afterwards asymptotically approaches the equilibrium position. In this case, the overdamped oscillator solution \eqref{ods3} also has a zero at
\begin{equation}
t_{od}=\frac{1}{2\alpha}\ln\frac{(\gamma-\alpha)+|v_0|}{(\gamma+\alpha)-|v_0|}\,.
\label{tod}
\end{equation}
Depending on the magnitude of $\gamma$, the overdamped oscillator may, or may not, overshoot the equilibrium position. We can choose an overdamped $\gamma$ such that the factor multiplying the first exponential in the parenthesis in \eqref{ods3} cancels, which happens when
\begin{equation}
\tilde{\gamma}=\frac{\left(\frac{|v_0|}{x_0}\right)^2+\omega_0^2}{2\left(\frac{|v_0|}{x_0}\right)} \,.
\label{gammatildaS}
\end{equation}
The solution to the overdamped oscillator then simplifies into
\begin{equation}
x_{od}(t)=\frac{|v_0|-(\tilde{\gamma}-\tilde{\alpha})x_0}{2\tilde{\alpha}}e^{-(\tilde{\gamma}+\tilde{\alpha}) t}=x_0e^{-(\tilde{\gamma}+\tilde{\alpha}) t}\,,
\label{ods33}
\end{equation}   
where $\tilde{\alpha}=\sqrt{\tilde{\gamma}^2-\omega_0^2}$. For the overdamped coefficient $\tilde{\gamma}$, the denominator of the fraction under the logarithm in \eqref{tod} also vanishes, which means that the zero of the overdamped solution happens when $t_{od}\to\infty$. Thus, for $\gamma\in \left\langle\omega_0, \tilde{\gamma}\right\rangle$ the overdamped solution overshoots the equilibrium position once, and for $\gamma\in\left[\tilde{\gamma}, \infty\right\rangle$ it approaches the equilibrium position without ever overshooting it. Since $\tilde{\gamma}>\omega_0$ and $\tilde{\alpha}>0$ in \eqref{ods33}, it's clear that the overdamped oscillator \eqref{ods33} approaches the equilibrium faster than the critically damped oscillator. 

\begin{figure}[h!]
\begin{center}
\begin{tikzpicture}
        \begin{axis}[
        width=0.485\textwidth,
        height=0.3\textwidth,
        xmin = 0,
        xmax = 12,
        ymin = -21,
        ymax = 1,
        xtick={0,1,...,11,12},
        ytick={-20,-15,...,-5,0},
        every tick label/.append style={font=\small},
        ylabel near ticks,
        xlabel near ticks,
        xlabel = {\small $t\,[\omega_0^{-1}]$},
        ylabel = {\small $\log_{10} \left(E(t)/E_0\right)$},
        legend entries = {\scriptsize $\gamma=0.9\omega_0$\\\scriptsize $\gamma=\omega_0$\\\scriptsize $\gamma=1.245\omega_0$\\\scriptsize $\gamma=1.25\omega_0$\\\scriptsize $\gamma=1.255\omega_0$\\},
        legend style={at={(0.02,0.02)},anchor=south west}
        ]
        \addplot [domain=0:12,samples=200,thick, red](\x,{log10(((0.28*exp(-1.8*\x)*(5.261+4.261*cos(deg(0.872*\x+2.387))+2.064*sin(deg(0.872*\x+2.387)))))});
        \addplot [domain=0:12,samples=200,thick, dashed](\x,{log10(0.2*exp(-2*\x)*(5+2*\x*(\x-3)))});
        \addplot [domain=0:12,samples=300,thick, purple](\x,{-1.08139*\x+log10(-0.0072724+1.00727*cosh(1.48327*\x)-1.00723*sinh(1.48327*\x))});
        \addplot [domain=0:12,samples=200,thick, densely dotted, blue](\x,{-1.737*\x});
        \addplot [domain=0:12,samples=300,thick, densely dashed,cyan](\x,{-1.09008*\x+log10(0.0069562+0.993041*cosh(1.51661*\x)-0.993001*sinh(1.51661*\x))});
        \addplot [domain=0:12,samples=100,dashed,black](\x,{-6.8});
        \node[] at (axis cs: 8,-1.7) {(a)};
        \end{axis}  
    \end{tikzpicture}
\end{center}
\begin{center}
\begin{tikzpicture}
        \begin{axis}[
        width=0.485\textwidth,
        height=0.3\textwidth,
        xmin = 0,
        xmax = 7,
        ymin = -0.4,
        ymax = 1.1,
        xtick={0,1,...,6,7},
        ytick={-0.4,-0.2,...,0.8,1.0},
        every tick label/.append style={font=\small},
        ylabel near ticks,
        xlabel near ticks,
        xlabel = {\small $t\,[\omega_0^{-1}]$},
        ylabel = {\small $x(t)/x_0$},
        legend entries = {\scriptsize $\gamma=0.9\omega_0$\\\scriptsize $\gamma=\omega_0$\\\scriptsize $\gamma=1.25\omega_0$\\}
        ]
        \addplot [domain=0:7,samples=100,thick,red](\x,{2.713*exp(-0.9*\x)*cos(deg(0.436*\x+1.1934))});
        \addplot [domain=0:7,samples=100,thick,dashed](\x,{(1-\x)*exp(-\x)});
        \addplot [domain=0:7,samples=100,thick,blue, densely dotted](\x,{exp(-1.25*\x)*(cosh(0.75*\x)-sinh(0.75*\x))});
        \addplot [domain=0:7,samples=50,dashed,black](\x,{0});
        \node[] at (axis cs: 4.5,0.9) {(b)};

        \end{axis}       
    \end{tikzpicture}
\end{center}
\caption{(a) The base 10 logarithm of the ratio (9) for the underdamped oscillator with $\gamma=0.9\omega_0$ (solid red curve), the critically damped oscillator (black dashed curve), and the overdamped oscillators with $\gamma=1.245\omega_0$ (purple curve), $\gamma=1.25\omega_0$ (blue dotted curve) and $\gamma=1.255\omega_0$ (cyan dashed curve). The curves are drawn with the condition $|v_0|=2\omega_0 x_0$. The energy level $E=10^{-7}E_0$ is shown for easier comparison of curves. (b) The displacements of the oscillators as a function of time.}
\label{fig:Fig11}
\end{figure}
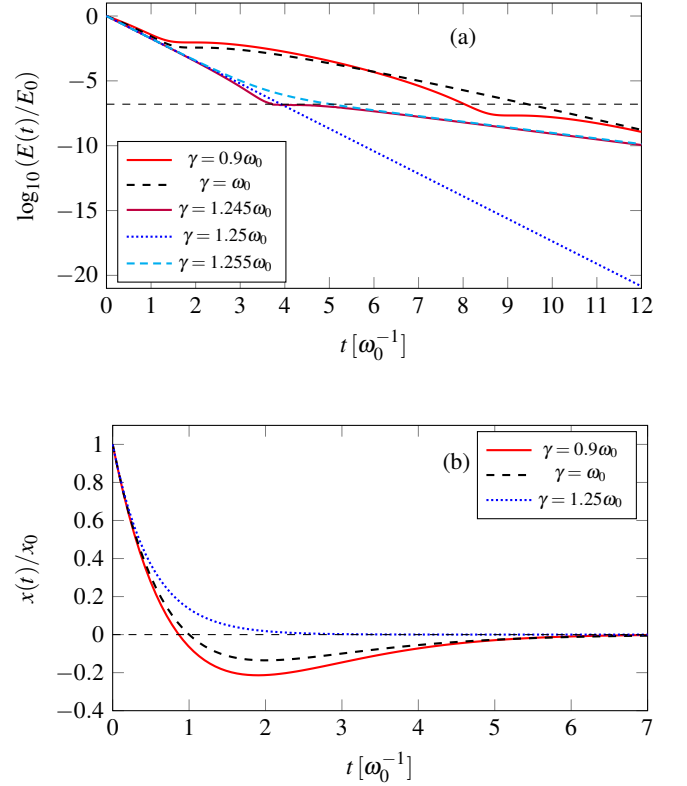

In Fig.\,\ref{fig:Fig11}(a) we show the base 10 logarithm of the ratio (9) for the underdamped oscillator with $\gamma=0.9\omega_0$ as the solid red curve, the critically damped oscillator as the black dashed curve, and the overdamped oscillators with $\gamma=1.245\omega_0$ as the purple curve, $\gamma=1.25\omega_0$ as the blue dotted curve and $\gamma=1.255\omega_0$ as the cyan dashed curve. To draw the curves we have selected the condition $|v_0|=2\omega_0 x_0$. With this choice, it follows from \eqref{gammatildaS} that $\tilde{\gamma}=1.25\omega_0$, which is shown as the blue dotted curve in Fig.\,\ref{fig:Fig11}(b). We can clearly see that the overdamped oscillator with $\tilde{\gamma}$ decays faster than both the underdamped and the critically damped oscillators. For values of the damping coefficient that are slightly smaller than $\tilde{\gamma}$, like $\gamma=1.245\omega_0$ shown with the purple curve, the system could get somewhat sooner to some desired energy level, which can be seen for the energy level of $10^{-7}E_0$ in this example. We could, in principle, optimise the damping coefficient of the overdamped oscillator, but, as Fig.\,\ref{fig:Fig11}(a) suggests, this would bring an insignificant time benefit compared to $\tilde{\gamma}$. Thus, we conclude that for this type of initial conditions, the fastest route to equilibrium is achieved by simply choosing the parameters of the system so that the damping coefficient is as close to $\tilde{\gamma}$ as possible. 

\section{Critically damped energy to underdamped energy ratio for initial conditions $x_0>0$ and $v_0=0$}

In Fig.\,\ref{fig:Fig4}(a) we show the ratio of the energy of a critically damped oscillator and the energy of an underdamped oscillator
\begin{equation}
\frac{E_c(t)}{E_{ud}(t)}=\frac{\dot x_c(t)^2+\omega_0^2x_c(t)^2}{\dot x_{ud}(t)^2+\omega_0^2x_{ud}(t)^2},
\label{EcEud}
\end{equation}
for the case when $\gamma=0.9\omega_0$. The maxima of the ratio \eqref{EcEud} are also the maxima of the difference
\begin{equation}
\ln\frac{E_c(t)}{E_{ud}(t)} = \ln\frac{E_c(t)}{E_0}-\ln\frac{E_{ud}(t)}{E_0}\,,
\label{energdiff}
\end{equation}
and thus, one can find the times $\tau_{max_n}$ for which the $n$-th maximum of the ratio \eqref{EcEud} is achieved, from the condition
\begin{equation}
\frac{\textrm{d}}{\textrm{d}t}\left(\ln\frac{E_c(t)}{E_0}-\ln\frac{E_{ud}(t)}{E_0}\right)\bigg|_{t=\tau_{max_n}}=0\,.
\label{maxcon}
\end{equation}
This means that the maxima are achieved at moments in time for which the relative energy losses of the critically damped and the underdamped oscillators are of equal values (for clarification of terminology, please see equations (11) and (13), as well as the paragraph following them). In Fig.\,\ref{fig:Fig4}(b) we show the energy loss rate of the critically damped oscillator as the black dashed curve and the energy loss rate of the underdamped oscillator for the case when $\gamma=0.9\omega_0$ as the red solid curve. We see that the three maxima of the ratio \eqref{EcEud}, shown in Fig.\,\ref{fig:Fig4}(a), happen at the first, the third, and the fifth crossing of the two energy loss rates, indicated by arrows and denoted by $\tau_{max_n}$, with $n=\left\lbrace 1, 2, 3\right\rbrace$. The values of $\tau_{max_n}$ are obtained numerically form the graph, since they can not be determined analytically. 
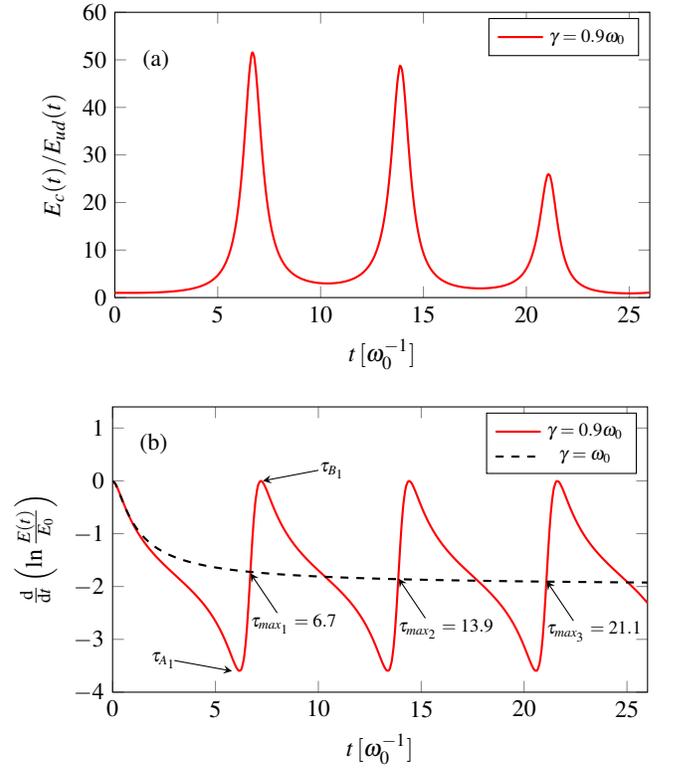
\begin{figure}
\begin{center}
\begin{tikzpicture}
        \begin{axis}[
        width=0.485\textwidth,
        height=0.3\textwidth,
        xmin = 0,
        xmax = 26,
        ymin = 0,
        ymax = 60,
        xtick={0,5,...,20,25},
        ytick={0,10,...,50,60},
        every tick label/.append style={font=\small},
        ylabel near ticks,
        xlabel near ticks,
        xlabel = {\small $t\,[\omega_0^{-1}]$},
        ylabel = {\small $E_c(t)/E_{ud}(t)$},
        legend entries = {\scriptsize $\gamma=0.9\omega_0$\\}
        ]
        \addplot [smooth, domain=0:26,samples=200,thick, red](\x,{0.19*exp(-0.2*\x)*((1+\x)^2+\x^2)/(sin(deg(0.436*\x))^2+cos(deg(0.436*\x-1.119))^2)});

        \node[] at (axis cs: 2,50) {(a)};
        
        \end{axis}  
    \end{tikzpicture}
\end{center}
\begin{center}
\hskip -4mm
\begin{tikzpicture}
        \begin{axis}[
        width=0.485\textwidth,
        height=0.3\textwidth,
        xmin = 0,
        xmax = 26,
        ymin = -4,
        ymax = 1.4,
        xtick={0,5,...,20,25},
        ytick={-4,-3,...,0,1},
        every tick label/.append style={font=\small},
        ylabel near ticks,
        xlabel near ticks,
        xlabel = {\small $t\,[\omega_0^{-1}]$},
        ylabel = {\small $\frac{\textrm{d}}{\textrm{d}t}\left(\ln\frac{E(t)}{E_0}\right)$},
        legend entries = {\scriptsize $\gamma=0.9\omega_0$\\\scriptsize $\gamma=\omega_0$\\}
        ]
        \addplot [smooth,domain=0:26,samples=200,thick, red](\x,{-1.8+0.872*(sin(deg(0.436*\x))*cos(deg(0.436*\x))-cos(deg(0.436*\x-1.119))*sin(deg(0.436*\x-1.119)))/(sin(deg(0.436*\x))^2+cos(deg(0.436*\x-1.119))^2)});
        \addplot [domain=0:26,samples=200,thick, dashed](\x,{-2+2*(1+2*\x)/((1+\x)^2+\x^2)});

        \node[] at (axis cs: 2,0.7) {(b)};
        
        \draw[->,>=stealth] (axis cs: 8,-2.5) -- (axis cs: 6.7,-1.75);
        \node[] at (axis cs: 8.8,-2.7) {\scriptsize $\tau_{max_1}=6.7$};
        \draw[->,>=stealth] (axis cs: 15.2,-2.6) -- (axis cs: 13.9,-1.85);
        \node[] at (axis cs: 16.2,-2.8) {\scriptsize $\tau_{max_2}=13.9$};
        \draw[->,>=stealth] (axis cs: 22.4,-2.65) -- (axis cs: 21.1,-1.9);
        \node[] at (axis cs: 23.4,-2.85) {\scriptsize $\tau_{max_3}=21.1$};
        
        \draw[->,>=stealth] (axis cs: 3,-3.4) -- (axis cs: 5.8,-3.6);
        \node[] at (axis cs: 2.5,-3.4) {\scriptsize $\tau_{A_1}$};
        \draw[->,>=stealth] (axis cs: 10,0.2) -- (axis cs: 7.3,0);
        \node[] at (axis cs: 10.7,0.2) {\scriptsize $\tau_{B_1}$};
        
        \end{axis}  
    \end{tikzpicture}
\end{center}
\caption{(a) The ratio of the energy of a critically damped oscillator and the energy of an underdamped oscillator for the case when $\gamma=0.9\omega_0$. The maxima are achieved at moments for which the energy loss rates of the critically damped and the underdamped oscillators are of equal values. (b) The energy loss rates of the critically damped oscillator (black dashed curve) and the underdamped oscillator for the case when $\gamma=0.9\omega_0$ (red solid curve). The three maxima shown in (a) happen at moments indicated by arrows and denoted by $\tau_{max_n}$, with $n=\left\lbrace 1, 2, 3\right\rbrace$. $\beta_{ud}$ equals 1 at a time instant denoted with $\tau_{A_1}$ and equals 0 at a time instant denoted by $\tau_{B_1}$.}
\label{fig:Fig4S}
\end{figure}

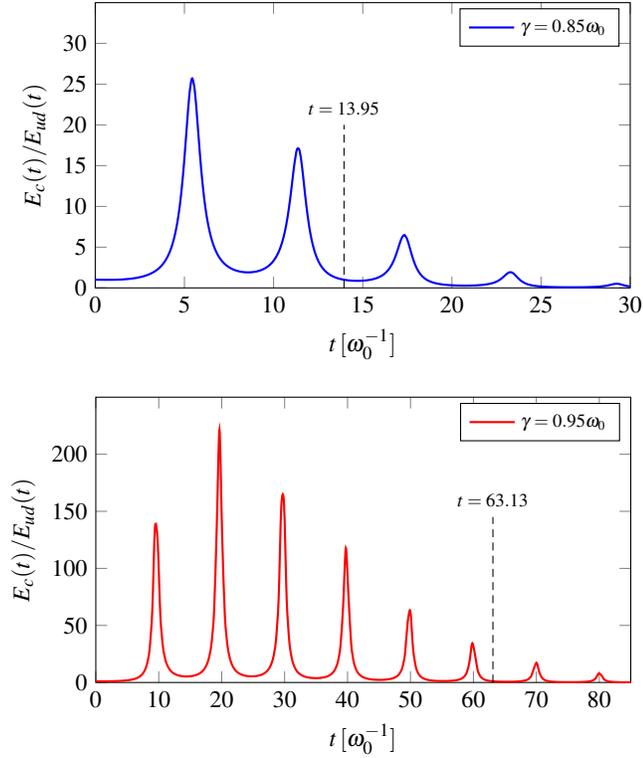
\begin{figure}[h!t!]
\begin{center}
\begin{tikzpicture}
        \begin{axis}[
        width=0.485\textwidth,
        height=0.3\textwidth,
        xmin = 0,
        xmax = 30,
        ymin = 0,
        ymax = 35,
        xtick={0,5,...,25,30},
        ytick={0,5,...,25,30},
        every tick label/.append style={font=\small},
        ylabel near ticks,
        xlabel near ticks,
        xlabel = {\small $t\,[\omega_0^{-1}]$},
        ylabel = {\small $E_c(t)/E_{ud}(t)$},
        legend entries = {\scriptsize $\gamma=0.85\omega_0$\\}
        ]
        \addplot [smooth, domain=0:30,samples=200,thick, blue](\x,{0.2775*exp(-0.3*\x)*((1+\x)^2+\x^2)/(sin(deg(0.527*\x))^2+cos(deg(0.527*\x-1.016))^2)});

        \draw[densely dashed] (axis cs: 13.95,0) -- (axis cs: 13.95,20);
        \node[] at (axis cs: 13.95,22) {\scriptsize $t=13.95$};
        
        \end{axis}  
    \end{tikzpicture}
\end{center}
\begin{center}
\hskip -4mm
\begin{tikzpicture}
        \begin{axis}[
        width=0.485\textwidth,
        height=0.3\textwidth,
        xmin = 0,
        xmax = 85,
        ymin = 0,
        ymax = 250,
        xtick={0,10,...,70,80},
        ytick={0,50,...,150,200},
        every tick label/.append style={font=\small},
        ylabel near ticks,
        xlabel near ticks,
        xlabel = {\small $t\,[\omega_0^{-1}]$},
        ylabel = {\small $E_c(t)/E_{ud}(t)$},
        legend entries = {\scriptsize $\gamma=0.95\omega_0$\\}
        ]
        \addplot [smooth,domain=0:85,samples=200,thick, red](\x,{0.0975*exp(-0.1*\x)*((1+\x)^2+\x^2)/(sin(deg(0.312*\x))^2+cos(deg(0.312*\x-1.253))^2)});

        \draw[densely dashed] (axis cs: 63.1,0) -- (axis cs: 63.1,145);
        \node[] at (axis cs: 63.1,160) {\scriptsize $t=63.13$};
        
        \end{axis}  
    \end{tikzpicture}
\end{center}
\caption{The ratio of the energy of a critically damped oscillator and the energy of an underdamped oscillator for the cases when $\gamma=0.85\omega_0$ and $0.95\omega_0$. The dashed vertical lines denote the first moment in which the energy of the critically damped oscillator becomes equal to the energy or the underdamped oscillator.}
\label{fig:Fig5}
\end{figure}

For the chosen case of the underdamped oscillator the maximal magnitude of the energy loss rate equals $4\gamma=3.6\omega_0$. This means that, according to (13), $\beta_{ud}=1$ at a time instant we denote with $\tau_{A_1}$. In this moment, the system arrives for the first time to the equilibrium position and its total energy is its kinetic energy. In contrast, the energy loss rate is zero when $\beta_{ud}=0$ at a time instant denoted by $\tau_{B_1}$, in which the total energy of the oscillator is equal to its potential energy. To be clear, here we chose an underdamped oscillator for which $\gamma=0.9\omega_0$ only to set an example. Choosing any  other $\gamma$ in the underdamped regime yields qualitatively the same behavior. The quantitative differences for various cases are in the number of maxima before the equality of the energies of the underdamped and the critically damped oscillators is achieved, and in the height of these maxima. In Fig.\,\ref{fig:Fig5}(a) and (b) we show the ratio \eqref{EcEud} for $\gamma=0.85\omega_0$ and $0.95\omega_0$. In the same figure, we use dashed vertical lines to denote the first moment in which the energy of the critically damped oscillator becomes equal to the energy or the underdamped oscillator.

Here we comment on an approximation for the moments $\tau_{max_n}$ for which the ratio \eqref{EcEud} has maxima. Visual inspection of Fig. \,\ref{fig:Fig4S} suggests that a good first guess for the moments $\tau_{max_n}$ is in the vicinity of moments when $\beta_{ud}=1/2$. Specifically, for the case shown in Fig. \,\ref{fig:Fig4S} we find $\beta_{ud}(\tau_{max_1}) = 0.477$, and similarly $\beta_{ud}(\tau_{max_{2,3}}) \approx 0.5$. Hence, the moments in time when maxima are achieved can be excellently approximated by the moments $t_n$ for which $\beta_{ud}(t_n)=1/2$, i.e. when the kinetic energy of the oscillator is equal to its potential energy. This condition can be written down analytically as
\begin{equation}
\dot x_{ud}(t_n)^2=\omega_0^2x_{ud}(t_n)^2.
\label{condition_t1}
\end{equation}
When taking the square root of \eqref{condition_t1} one must take into account that at times $\tau_{max_n}$ (the first, the third and the fifth crossing of energy loss rates in Fig. 4(b)) the underdamped oscillator is moving away from the equilibrium position, meaning that the signs of the velocity and the elongation are the same. Then we simply have
\begin{equation}
\dot x_{ud}(t_n)=\omega_0 x_{ud}(t_n).
\label{condition_t2}
\end{equation}
Solving \eqref{condition_t2} we get
\begin{equation}
t_n=\frac{1}{\omega}\left(\arctan\left(-\frac{\omega}{\omega_0+\gamma}\right)+n\pi\right),
\label{t}
\end{equation}
where $n\in\mathbb{N}$. We are interested in the total number of maxima that occur before the first crossing of the energies of the underdamped oscillator and the critically damped oscillator. In the case when $\gamma=0.9\omega_0$ we only take $n=\lbrace1,2,3\rbrace$ since these are the only crossings that occur before $t=24.24\omega_0^{-1}$. In Table SII. we give the results for the moment $\tau_{max_1}$ obtained numerically and for $t_1$ obtained from \eqref{t} with $n=1$, with results  rounded to two decimal points, for five damping coefficients in the interval $\gamma\in\left[0.7\omega_0, \omega_0\right\rangle$. It can easily be seen that $\tau_{max_1}\approx t_1$ with relative error less than $1\%$. The calculated approximation holds very well for other maxima as well, i.e., $\tau_{max_n}\approx t_n$ for $n>1$ in the same interval of damping coefficients. 
\begingroup
\setlength{\tabcolsep}{5.2pt} 
\renewcommand{\arraystretch}{1.5} 
\begin{table}[h!t!]
\begin{tabular}{|c|c|c|c|} 
 \hline
 $\gamma [\omega_0]$ & $\tau_{max_1} [ \omega_0^{-1}]$ & $t_1 [ \omega_0^{-1}]$ & $\left(\tau_{max_1} - t_1 \right)/\tau_{max_1} [ \% ]$ \\ 
 \thickhline
 $0.7$ &  $3.81$  & $3.84$ & $-0.79$ \\ 
 \hline
 $0.8$ &  $4.70$  & $4.70$ & $0$ \\ 
 \hline
 $0.9$ &  $6.70$  & $6.69$ & $0.15$ \\ 
 \hline
 $0.95$ & $9.56$ & $9.55$ & $0.1$ \\ 
 \hline
 $0.99$ & $21.78$  & $21.77$ & $0.05$ \\ 
 \hline 
\end{tabular}
\caption{The numerically obtained moment $\tau_{max_1}$ (col.2) and $t_1$ obtained from \eqref{t} with $n=1$ (col.3) for five different damping coefficients (col.1). It can easily be seen that $\tau_{max_1}\approx t_1$ with relative error less than $1\%$ (col.4).}
\end{table}
\endgroup

\section{Optimisation of the underdamped $\gamma$ with respect to $\tau_{A_2}$}

\begingroup
\setlength{\tabcolsep}{5.2pt} 
\renewcommand{\arraystretch}{1.5} 
\begin{table}[h!t!]
\begin{tabular}{ |c|c|c|c|c| } 
 \hline
 $E(t)/E_0$ & $\gamma_2 \left[ \omega_0 \right]$ & $\tau_{A_2} [ \omega_0^{-1}]$ & $\tau_{c}[ \omega_0^{-1}]$ & $\left(\tau_{c} - \tau_{A_2} \right)/\tau_{c} \left[ \% \right]$ \\ 
 \thickhline
 $10^{-4}$ & \ $0.6477$ \ & $7.11$ & $6.96$ & $-2.16$ \\ 
 \hline
 $10^{-6}$ & \ $0.7767$ \ & $8.89$ & $9.56$ & $7$ \\ 
 \hline
 $10^{-8}$ & \ $0.8492$ \ & $10.84$ & $12.10$ & $10.41$ \\ 
 \hline
 $10^{-10}$ & \ $0.8926$ \ & $12.90$ & $14.57$ & $11.46$ \\ 
 \hline
 $10^{-12}$ & \ $0.9201$ \ & $15.01$ & $17.03$ & $11.86$ \\ 
 \hline
 $10^{-14}$ & \ $0.9385$ \ & $17.18$ & $19.46$ & $11.72$ \\ 
 \hline
 $10^{-16}$ & \ $0.9513$ \ & $19.37$ & $21.88$ & $11.47$ \\ 
 \hline
 $10^{-18}$ & \ $0.9605$ \ & $21.57$ & $24.28$ & $11.16$ \\ 
 \hline
\end{tabular}
\caption{The underdamped coefficients $\gamma_2$ (col.2) obtained from condition \eqref{condition5} for various choices of $\delta$ (col.1). The last three columns show the moments in time when the condition \eqref{condition5} is satisfied for the $\gamma_2$ (col.3), the time when the energy of a critically damped oscillator drops to the same level (col.4) and the relative time advantage of the underdamped oscillator compared to the critically damped oscillator (col.5).}
\end{table}
\endgroup

\hskip -3mm
Using (20) with $n=2$ we get  
\begin{equation}
\exp\left[-2\frac{\gamma}{\sqrt{\omega_0^2-\gamma^2}}\left(\frac{3\pi}{2}+\arctan\frac{\gamma}{\sqrt{\omega_0^2-\gamma^2}}\right)\right]=10^{-\delta}\,.
\label{condition4}
\end{equation}
Introducing the shorthand notation $f(\gamma)$ for the left-hand side of \eqref{condition4}, we can determine the gamma that reaches the desired energy level at $\tau_{A_2}$ graphically by using the condition
\begin{equation}
\log_{10}(f(\gamma))=-\delta\,.
\label{condition5}
\end{equation}
In Table SIII. we show the results obtained this way for various choices of $\delta$. We can see that the relative advantage of the underdamped oscillator compared to the critically damped oscillator, given in column five is smaller than was the case in Table I. This is expected, since the energy of the underdamped oscillator is closer in value to the energy of the critically damped oscillator after each cycle, until it eventually surpasses it (see Fig.\,3(a) and (b)). It is important to note the negative result in the first row of column five, indicating that the critically damped oscillator reaches the energy $E(t)=10^{-4}E_0$ before the underdamped oscillator. We could, of course, also look for an damping coefficient such that we declare that the underdamped oscillator has settled down in the equilibrium at the third, or fourth, etc., pass through the equilibrium position, but we would get ever smaller relative advantage over the critical oscillator. 

\section{Experiment with a detectable overshoot}

\begin{figure}[h!t!]
\centering
\begin{tikzpicture}
\node[above right] (img) at (0,0) {\includegraphics[width=0.45\textwidth]{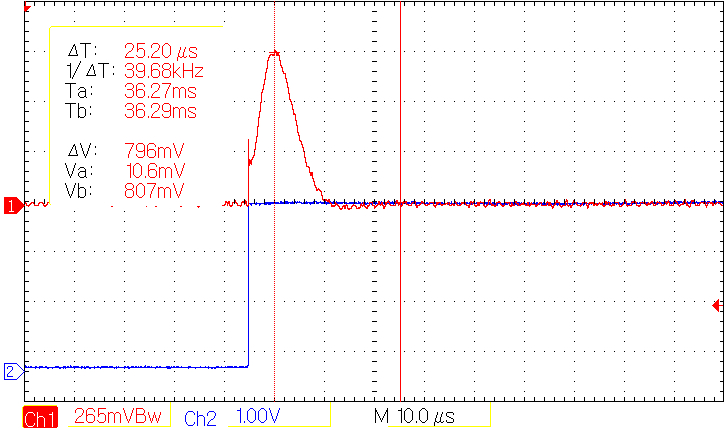}};
\node at (190pt,120pt) {\large $R=5000\,\Omega$};
\end{tikzpicture}
\caption{The readout on the oscilloscope in the case of the underdamped circuit with $R=5\,\textrm{k}\Omega$ (red curve). The time interval required for the voltage to drop from the maximal value to the resolution of the oscilloscope was measured to be $(25.2 \pm 0.4)$\,$\mu$s. An overshoot of the signal under the reference voltage can clearly be seen.}
\label{fig:overshoot}
\end{figure}

As an example of the experiment with a detectable overshoot, we give the readout of the voltage for a resistance value of 5\,k$\Omega$ on Fig.\ref{fig:overshoot}, where we noticed that the second crossing happened at $(25.2 \pm 0.4)$\,$\mu$s, after which the voltage was below the oscilloscope resolution. 

\end{document}